\documentclass[preprint,12pt]{elsarticle}

\usepackage{amssymb}
\usepackage{mathtools}
\usepackage{amsmath}
\usepackage{tikz}
\usetikzlibrary{shapes,decorations,patterns}
\usetikzlibrary{arrows, snakes, decorations.pathmorphing}
\usepackage{bm}
\usepackage{lineno}
\usepackage{fancyhdr}
\usepackage[T1]{fontenc}

\newcommand{\mynormaltext}[1]{{{#1}}}

\journal{International Journal of Heat and Mass Transfer}
\date{December 10, 2021}

\fancypagestyle{firstpage}
{
	
	\fancyhead[L]{\scriptsize \copyright 2022. This manuscript version is made available under the CC-BY-NC-ND 4.0 license https://creativecommons.org/licenses/by-nc-nd/4.0/
	}    
	\fancyhead[R]{}
	\fancyfoot[L]{\scriptsize Preprint submitted to International Journal of Heat and Mass Transfer}    
	\fancyfoot[R]{\scriptsize December 10, 2021}

}

\begin{document}

\begin{frontmatter}

\title{Evaporation of volatile droplets subjected to flame-like conditions}

\author[bsc]{Ambrus Both\corref{corAB}}
\ead{ambrus.both@bsc.es}
\cortext[corAB]{Corresponding author}

\author[bsc]{Daniel Mira}

\author[bsc]{Oriol Lehmkuhl}

\affiliation[bsc]{organization={Barcelona Supercomputing Center (BSC)},
            addressline={Pla\c{c}a Eusebi G\"uell 1-3},
            city={Barcelona},
            postcode={08034}, 
            country={Spain}}

\begin{abstract}
This work assesses Lagrangian droplet evaporation models frequently used in spray combustion simulations, with the purpose of identifying the influence of modeling decisions on the single droplet behavior. 
Besides more simplistic models, the evaluated strategies include 
a simple method to incorporate Stefan flow effects in the heat transfer (Bird's correction),
a method to consider the interaction of Stefan flow with the heat and mass transfer films (Abramzon-Sirignano model),
and a method to incorporate non-equilibrium thermodynamics (Langmuir-Knudsen model).
The importance of each phenomena is quantified analytically and numerically under various conditions.
Evaporation models ignoring Stefan flow are found to be invalid under the studied conditions.
The Langmuir-Knudsen model is also deemed inadequate for high temperature evaporation, while \mynormaltext{Bird's correction} and the Abramzon-Sirignano model \mynormaltext{are} identified as the most relevant for numerical studies of spray combustion systems. \mynormaltext{Latter is the most elaborate model studied here, as it considers Reynolds number effects beyond the empirical correlation of Ranz and Marshall derived for low-transfer rates. Thus, the Abramzon-Sirignano model is identified as the state of the art alternative in the scope of this study. }
\end{abstract}


\end{frontmatter}

\thispagestyle{firstpage}

\section{Introduction}

Spray combustion simulations overwhelmingly use an Eulerian-Lagrangian approach to account for the gas and liquid phase respectively. \cite{sazhin2014droplets}
\mynormaltext{In this approach, the} liquid droplets are represented by point particles, 
that move independently in the \mynormaltext{computational} %
domain interacting with the gas phase.
This modeling strategy \mynormaltext{is known to be valid} %
in the dilute spray regime, where the liquid volume fraction is below 0.001. \cite{jenny2012modeling}
A crucial aspect besides the kinematic modeling of these computational particles, is the heat and mass transfer process resulting in the evaporation of the fuel that ultimately feeds \mynormaltext{the reacting front in} combustion \mynormaltext{simulations}. 

\mynormaltext{Various strategies have been developed %
to account for droplet evaporation 
considering different aspects} 
of heat and mass transfer.
Miller~et~al.~\cite{miller1998evaluation} introduced a unified framework of 
different evaporation models,
and conducted a comparative study.
However,  
as the models were developed under easily measurable conditions, 
corresponding to rather large droplets ($\sim 1 \ \mathrm{mm}$), their direct application to spray combustion could be questionable under certain conditions\mynormaltext{, where droplet diameters are rather small ($\sim 1 \ \mathrm{\mu m} .. 10 \ \mathrm{\mu m}$). 
The main issue is the homogeneity in the interior of these droplets. 
While, in measurements of $1 \ \mathrm{mm}$ droplets, temperature variations inside the droplet are important, in the range of interest these can be negligible.}
The topic enjoys renewed interest, prompted by the recent experimental investigation of Verdier~et~al.~\cite{verdier2017experimental} using Global Rainbow Thermometry to characterize the mean droplet temperatures in a complex lab-scale n-heptane spray flame.
This flame was numerically investigated in the Workshop on Measurement and Computation of Turbulent Spray Combustion by different groups \cite{both2017rans,noh2018comparison,sitte2019large, ausilio2019study, chatelier2020large, benajes2021analysis} using Lagrangian droplet models for the evaporating spray cloud. 
Specifically, Noh~et~al.~\cite{noh2018comparison} compared various evaporation models following Miller~et~al.~\cite{miller1998evaluation}, using large-eddy simulation (LES) to asses the droplet temperature predictions. 
These studies provide an \mynormaltext{overview} %
of the state of the art of spray combustion simulations of gas turbine model combustors, 
however the underlying behavior of the droplet evaporation models requires further studying.

The objectives of the present work are 
i) to clarify the definition of evaporation models from first principles using a film theory approach, 
ii) to provide further understanding on the behavior of the evaporation models under realistic flame-like conditions, 
iii) to quantify the error made by the models and the relative importance of specific models, 
and finally iv) to study the behavior of fuels characterized by different volatility. 
Latter aspect is evaluated by using OME1 (dimethoxymethane, formerly methylal), and 3 alkanes: n-heptane, n-decane, and n-dodecane, 
that have a boiling point of $315.0 \ \mathrm{K}$, $371.5 \ \mathrm{K}$, $447.5 \ \mathrm{K}$, and $489.5 \ \mathrm{K}$ respectively. 
In this aspect, OME1 is especially interesting, as it is a high volatility \mynormaltext{fuel %
showing} the distinctive effects of high evaporative mass flux even in moderate temperature environments.

In section~\ref{sec:modelling}, the models are presented \mynormaltext{and 
analyzed} under different seen gas conditions in subsections~\ref{subsec:wet_bulb} and~\ref{subsec:single_drop} in terms of equilibrium temperatures (wet-bulb conditions) and single droplet simulations respectively. 
Finally conclusions are drawn.

\section{Lagrangian droplet modeling}
\label{sec:modelling}

Fuel droplets are commonly modeled in CFD calculations, 
as stand-alone Lagrangian particles interacting with the gas phase. 
In this framework, the heat and mass transfer is usually treated 
as an exchange, between the practically infinite gas phase, 
and the spherical particle. 
These assumptions are justified in 
the dilute spray regime, 
\mynormaltext{where} the direct influence of droplet to droplet interactions is negligible. 
Furthermore, the length scale of the droplets in typical spray combustion systems 
is \mynormaltext{of} $\mathcal{O}\left(10 \ \mathrm{\mu m}\right)$, 
that is below the smallest length scales of thermo-chemical nonhomogenities associated to the flame thickness: $\mathcal{O}\left(100 \ \mathrm{\mu m}\right)$ \mynormaltext{\cite[§5.1.2]{poinsot2005theoretical}}, thus the far-field behavior of the gas phase may be regarded homogeneous with respect to the droplets.

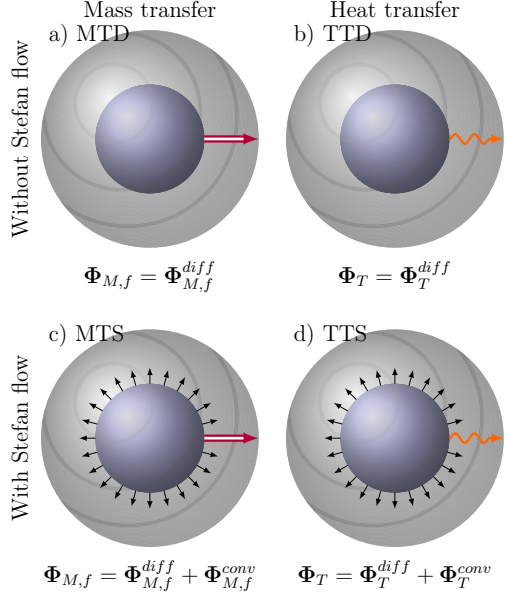
\begin{figure}[h!]
    \centering
    
    \resizebox{0.5\linewidth}{!}{
    \begin{tikzpicture}[
           heat/.style={decorate,decoration={snake,post length=1mm}, color=red!60!yellow, very thick},
           mass/.style={double  distance=1.4pt, color=black!30!blue!30!red, very thick}
       ]
       \node[above] at (0,2.1) {Mass transfer};
       \node[rotate=90,above] at (-2.1,0) {Without Stefan flow};

	    \shade[ball color = blue!30,
	    opacity = 1.0
	    ] (0,0,0) circle (1cm);

	    \shade[ball color = gray,
	    opacity = 0.2
	    ] (0,0,0) circle (2cm);

	    \draw [-latex, mass](1,0) --  (2,0);
	    
	    \node[right] at (-2,1.9) {a) MTD};
	    \node[below] at (0,-2.1) {$\bm{\Phi}_{M,f} =\bm{\Phi}_{M,f}^{diff}$};

	    \node[above] at (4.5,2.1) {Heat transfer};
	    \shade[ball color = blue!30,
	    opacity = 1.0
	    ] (4.5,0,0) circle (1cm);
	    
	    \shade[ball color = gray,
	    opacity = 0.2
	    ] (4.5,0,0) circle (2cm);
	    \draw [-latex, heat](5.5,0) --  (6.5,0);
	    
	    \node[right] at (2.5,1.9) {b) TTD};
	    \node[below] at (4.5,-2.1) {$\bm{\Phi}_{T} =\bm{\Phi}_{T}^{diff}$};

        \node[rotate=90,above] at (-2.1,-5.5) {With Stefan flow};

	    \shade[ball color = blue!30,
	    opacity = 1.0
	    ] (0,-5.5,0) circle (1cm);
	    
	    \shade[ball color = gray,
	    opacity = 0.2
	    ] (0,-5.5,0) circle (2cm);

	    \foreach \a in {15,30,...,359}
        \draw[-latex] (0,-5.5) +(\a:1) -- +(\a:1.3);

	    \draw [-latex, mass](0,-5.5) +(1,0) --  +(2,0);
	    
	    \node[right] at (-2,-3.6) {c) MTS};
	    \node[below] at (0,-7.6) {$\bm{\Phi}_{M,f} =\bm{\Phi}_{M,f}^{diff} + \bm{\Phi}_{M,f}^{conv}$};

	    \shade[ball color = blue!30,
	    opacity = 1.0
	    ] (4.5,-5.5,0) circle (1cm);
	    
	    \shade[ball color = gray,
	    opacity = 0.2
	    ] (4.5,-5.5,0) circle (2cm);

	    \foreach \a in {15,30,...,359}
       \draw[-latex] (4.5,-5.5) +(\a:1) -- +(\a:1.3);

	    \draw [-latex, heat] (4.5,-5.5) +(1,0) --  +(2,0);
	    
	    \node[right] at (2.5,-3.6) {d) TTS};
	    \node[below] at (4.5,-7.6) {$\bm{\Phi}_{T} =\bm{\Phi}_{T}^{diff} + \bm{\Phi}_{T}^{conv}$};

	\end{tikzpicture}
	}
	
    \caption{Summary of different droplet heat and mass transfer model problems: a) Mass Transfer solely due to Diffusion (MTD), b) Thermal Transfer solely due to Diffusion (TTD), c) Mass Transfer including the convective effect of Stefan flow (MTS), d) Thermal Transfer including the convective effect of Stefan flow (TTS).}
    \label{fig:eq_scenarios}
\end{figure}

Many widely applied models study the phenomenon of evaporation 
based on film theory. 
Film theory postulates that the differences between interface and bulk states 
diminish in a finite $\delta_M$ and $\delta_T$ thickness for the mass and thermal transfer respectively.
Figure~\ref{fig:eq_scenarios} summarizes four theoretical scenarios 
for the treatment of heat and mass transfer between a spherical particle 
and its surroundings. 
\mynormaltext{The illustrated scenarios include Mass Transfer (MTD) and Thermal Transfer (TTD) solely due to Diffusion, and Mass Transfer (MTS) and Thermal Transfer (TTS) including the convective effect of Stefan flow: the blowing effect of intense evaporation.}
Stefan flow is an important phenomena affecting rapidly evaporating droplets, as it obstructs heat transfer from the high temperature gas to the droplet interface.
In this work, both heat and mass transfer are studied under the quasi-steady state assumption,
postulating that the boundary layer surrounding the droplet reaches its steady conditions infinitely fast. 
\mynormaltext{ This relaxation is characterized by a time scale corresponding to unity Fourier number: $Fo = \tfrac{t \mathcal{D}_{t,m} }{d_p^2}$, where $t$ is the time scale, $\mathcal{D}_{t,m}=\mathcal{O}\left(10 \ \mathrm{mm^2/s}\right)$ is the thermal diffusivity in the mixture, and $d_p$ is the droplet diameter. 
The relaxation to the steady state profiles is two orders of magnitude faster, than the evaporation process itself, as discussed in subsection~\ref{subsec:single_drop}. }

\subsection{Quasi-steady heat and mass transfer around a sphere}
\label{subsec:quasi_steady_heat_and_mass}

\mynormaltext{
This subsection presents the modeling framework of heat and mass transfer, 
that is subsequently used in the definition of the evaporation models and their analysis.
}
\mynormaltext{The heat and mass fluxes are derived in an isolated manner, 
allowing the step-by-step construction of the evaporation models, and the detailed insight in their behavior.}

The diffusive mass flux \mynormaltext{across gas phase boundary layer surrounding the droplet} is proportional to the gradient of the volatile species. 
For a species $f$ with mass fraction $Y_f$ dissolved in a bath gas $b$ of mass fraction $Y_b=1-Y_f$
the Hirschfelder's law defines the diffusive mass flux as:
\begin{align}
\bm{\Phi}_{M,f}^{diff} =  -\rho_{m} \mathcal{D}_{m}  \mathbf{\nabla} Y_f,
\end{align}
where $\rho_{m}$ is the density of the mixture, and $\mathcal{D}_{m}$ is the mass diffusion coefficient of species $f$ in the mixture.
Furthermore, in case the gas mixture has a net mass flux $\bm{\Phi}_{M}$,
species $f$ and $b$ are also transported by convection:
\begin{align}
\bm{\Phi}_{M,f}^{conv} =  Y_f \bm{\Phi}_{M}.
\end{align}
This latter flux is the one related to Stefan flow, i.e.: the net mass flux caused by the vapor leaving the droplet surface. 
\mynormaltext{The convective mass flux is negligible at low evaporation rates, but it is relevant under flame-like conditions, especially for highly volatile fuels.}
In the evaporation of single component droplets, \mynormaltext{the net mass flux} is the mass flux of the volatile species: $\bm{\Phi}_{M} = \bm{\Phi}_{M,f}$, under the assumption that the bath gas is practically insoluble in the liquid droplet.

Similarly, the diffusive flux of heat is given by Fourier's law of heat conduction:
\begin{align}
\bm{\Phi}_T^{diff} =  - \lambda_{m} \mathbf{\nabla} T,
\end{align}
where $\lambda_m$ is the thermal conductivity in the gas mixture. 
However, to determine the convective heat transport, the net mass flux is used again \mynormaltext{creating a coupling between the heat and mass transfer}. 
The enthalpy of the volatile component is defined using a first order approximation using the definition of specific heat. 
Thus, the convective heat flux is: 
\begin{align}
 \bm{\Phi}_{T}^{conv} = 
c_{p,vap,m} \bm{\Phi}_{M,f}  \left(T-T_0\right),
\end{align} where $c_{p,vap,m}$ is the specific heat of the vapor of species $f$, and $T_0$ is an appropriately chosen  reference temperature.

The problems illustrated in Fig.\ref{fig:eq_scenarios} are rotationally  symmetric, thus only the radial components of fluxes are non-zero. 
Under the quasi-steady assumption, mass and energy conservation implies that the surface integral of the radial mass and thermal fluxes $\bm{\Phi}_{M,f,r}$ and $\bm{\Phi}_{T,r}$ are constant within the film on concentric spheres: 
\begin{align}
    \label{eq:radial_mass_transfer_rate}
    \dot{m}_r &= 4 r^2 \pi \bm{\Phi}_{M,f,r} = const.,\\
    \label{eq:radial_heat_transfer_rate}
    \dot{Q}_r &= 4 r^2 \pi \bm{\Phi}_{T,r} = const. 
\end{align}
Assuming $\rho_m \mathcal{D}_m = const.$ and $\lambda_m =  const.$, the above two equations form ODEs for the unknowns: $Y_f(r)$ and $T(r)$ respectively, with the boundary conditions:
\begin{align}
    \label{eq:boundary_condition_mass}
    Y_f(r_p) &= Y_{f,i}, &Y_f(r_{BL,M}) &= Y_{f,s}, \\
    \label{eq:boundary_condition_heat}
    T(r_p) &= T_p, &T(r_{BL,T}) &= T_s, 
\end{align}
where $r_p$ is the droplet radius, $r_{BL,M} = r_p + \delta_M$ and $r_{BL,T} = r_p + \delta_T$ are the outer film radii of mass and thermal transfer, $Y_{f,i}$ is the vapor mass fraction on the droplet interface, $Y_{f,s}$ is the seen vapor mass fraction (far-field), $T_p$ is the droplet temperature, and $T_s$ is the seen gas temperature. 
The solutions of \mynormaltext{Eq.~\eqref{eq:radial_mass_transfer_rate}-\eqref{eq:boundary_condition_heat}} are presented in Tab.~\ref{tab:solution_profiles}.
The temperature profile in the presence of Stefan flow \mynormaltext{(TTS)} is derived below, 
while the other three solutions are rather straightforward and can be found in the literature. \cite{bird1960transport}

\begin{table}[h!]
    \centering
    \begin{tabular}{r|c}
     MTD & 
     $Y_f^{MTD} = Y_{f,i} + \left(Y_{f,s}-Y_{f,i}\right) \dfrac{ \tfrac{1}{r} - \tfrac{1}{r_p} }{\tfrac{1}{r_{BL,M}} - \tfrac{1}{r_p} }$ \\
     MTS & $\dfrac{1-Y_f^{MTS}}{1-Y_{f,i}} = \left(\dfrac{1-Y_{f,s}}{1-Y_{f,i}}\right)^{\dfrac{\frac{1}{r_p}-\frac{1}{r}}{\frac{1}{r_p}-\frac{1}{r_{BL,M}}}}$\\
     TTD & $T^{TTD} = T_p + \left(T_s-T_p\right) \dfrac{ \tfrac{1}{r} - \tfrac{1}{r_p} }{\tfrac{1}{r_{BL,T}} - \tfrac{1}{r_p} }$ \\
     TTS & 
     $T^{TTS} = T_p + \left(T_s-T_p\right) \dfrac{e^{\tfrac{\Xi_1}{r}} - e^{\tfrac{\Xi_1}{r_p}} }{e^{\tfrac{\Xi_1}{r_{BL,T}}} - e^{\tfrac{\Xi_1}{r_p}} }$
    \end{tabular}
    \caption{Solution profiles in the four studied cases of droplet heat and mass transfer, with $\Xi_1 = \tfrac{c_{p,vap,m} }{c_{p,m} Le_m} \tfrac{1}{1/r_p - 1/r_{BL,M}} \ln\left(\tfrac{1-Y_{f,i}}{1-Y_{f,s}}\right) $, assuming constant gas phase properties.}
    \normalsize
    \label{tab:solution_profiles}
\end{table}

To the authors' knowledge, the Stefan flow effects on heat transfer in a spherically symmetric systems are often taken to be the same as those derived in Cartesian coordinates for a flat plate without separate derivations. \cite{miller1998evaluation,bird1960transport}
\mynormaltext{It is found, as shown} below, that the same correction factor (often known as Bird's \cite{bird1960transport} correction) 
is used in spherical and Cartesian coordinates. 

In the presence of Stefan flow, the radial energy flux is composed by the diffusive and convective fluxes:
\begin{align}
    \label{eq:TTS_heat_flux}
    \bm{\Phi}_{T,r}^{TTS} = 
    - \lambda_{m} \dfrac{\mathrm{d}T}{\mathrm{d}r} 
    +c_{p,vap,m} \bm{\Phi}_{M,f,r}(r)  \left(T-T_0\right).
\end{align}
Thus, energy conservation is described by the following ordinary differential equation:
\begin{align}
\label{eq:stefan_heat_ODE}
\dfrac{\mathrm{d}}{\mathrm{d}r} 
\left(
r^2 \dfrac{\mathrm{d}T}{\mathrm{d}r}
+ \Xi_1 T
+ \Xi_2
\right) = 0,
\end{align}
where $\Xi_1$ and $\Xi_2$ are constants using the solution of mass transfer in the presence of Stefan flow: $\Xi_1 = \tfrac{c_{p,vap,m} }{c_{p,m} Le_m} \tfrac{1}{1/r_p - 1/r_{BL,M}} \ln\left(\tfrac{1-Y_{f,i}}{1-Y_{f,s}}\right) $, and $\Xi_2 = - \tfrac{c_{p,vap,m} }{c_{p,m} Le_m} \tfrac{1}{1/r_p - 1/r_{BL,M}} \ln\left(\tfrac{1-Y_{f,i}}{1-Y_{f,s}}\right) T_0$, 
where $c_{p,m}$ is the specific heat of the gas mixture, and $Le_m = \tfrac{\lambda_m}{c_{p,m} \rho_m \mathcal{D}_m}$ is the mass based Lewis number of the vapor in the bath gas.
After performing the variable transformation: $\theta = \tfrac{T-T_p}{T_s-T_p}$, the temperature and its derivative are:
$T=T_p + \theta \left(T_s-T_p\right)$, and $\tfrac{\mathrm{d}T}{\mathrm{d}r} = \left(T_s-T_p\right) \tfrac{\mathrm{d}\theta}{\mathrm{d}r}$. Eq.~\eqref{eq:stefan_heat_ODE} can be written as:
\begin{align}
\label{eq:stefan_heat_ODE_scaled}
\dfrac{\mathrm{d}}{\mathrm{d}r} 
\left(
r^2 \dfrac{\mathrm{d}\theta}{\mathrm{d}r}
+ \Xi_1 \theta  
+ \Xi_2^*
\right) = 0,
\end{align}
where $\Xi_2^*$ is a constant. 
Eq.~\eqref{eq:stefan_heat_ODE_scaled} is a separable differential equation, with the general solution: $\theta = C_{1,T} e^{\tfrac{\Xi_1}{r}} + C_{2,T}$,
where $C_{1,T}$ and $C_{2,T}$ are integration constants subject to boundary conditions. 
By imposing the boundary conditions: $\theta(r_p) = 0$, $\theta(r_{BL,T}) = 1$,
one achieves: $C_{1,T} = \left( e^{\Xi_1/r_{BL,T}} - e^{\Xi_1/r_p}\right)^{-1}$ and $C_{2,T}=-C_{1,T} e^{\Xi_1/r_p}$, thus the temperature profile in steady state is $T^{TTS}$
as shown in Tab.~\ref{tab:solution_profiles} for the TTS case.

\begin{table}[h!]
    \centering
    \begin{tabular}{r|c}
     MTD & 
     $\dot{m}_r^{MTD} = \pi \rho_{m} \mathcal{D}_{m} d_p \left(Y_{f,i}-Y_{f,s}\right) \dfrac{2}{1-\tfrac{r_p}{r_{BL,M}}} $ \\
     MTS &
     $\dot{m}_r^{MTS} = \pi \rho_{m} \mathcal{D}_{m} d_p \ln\left(\tfrac{1-Y_{f,s}}{1-Y_{f,i}}\right) \dfrac{2}{1-\tfrac{r_p}{r_{BL,M}}} $ \\
     TTD &
     $\dot{Q}_r^{TTD} =  \pi \lambda_{m} d_p \left(T_p-T_s\right) \dfrac{2}{1-\tfrac{r_p}{r_{BL,T}}}$\\
     TTS &
     $\dot{Q}_r^{TTS} =  \pi \lambda_{m} d_p \left(T_p-T_s\right) \dfrac{2 \dfrac{\Xi_1}{r_p} }{1-e^{\tfrac{\Xi_1}{r_{BL,T}}-\tfrac{\Xi_1}{r_p}}}$
    \end{tabular}
    \caption{Radial heat and mass flow rates with and without Stefan flow, with $\Xi_1 = \tfrac{c_{p,vap,m} }{c_{p,m} Le_m} \tfrac{1}{1/r_p - 1/r_{BL,M}} \ln\left(\tfrac{1-Y_{f,i}}{1-Y_{f,s}}\right) $, assuming constant gas phase properties}
    \normalsize
    \label{tab:flows}
\end{table}

The radial heat flux is given by substitution to Eq.\eqref{eq:TTS_heat_flux}, and can be expressed using $T_0=T_p$:
\begin{align}
\label{eq:FilmRHeatFluxExpr}
\bm{\Phi}_{T,r}^{TTS} &= 
-\lambda_{m} \left( \dfrac{\mathrm{d}T}{\mathrm{d}r} + \dfrac{\Xi_1}{r^2} \left(T-T_0\right)  \right) 
=
-\lambda_{m} 
\left(T_s-T_p\right) \dfrac{1}{r^2} \dfrac{ -\Xi_1  e^{\tfrac{\Xi_1}{r_p}} }{e^{\tfrac{\Xi_1}{r_{BL,T}}} - e^{\tfrac{\Xi_1}{r_p}} }.
\end{align}
Finally the heat transfer rate is given by Eq.\eqref{eq:radial_heat_transfer_rate}:
\begin{align}
\label{eq:TTS_heat_transfer_rate}
\dot{Q}_r^{TTS} =  \pi \lambda_{m} d_p \left(T_p-T_s\right) \left. \left(2 \dfrac{\Xi_1}{r_p} \right) \middle/ \left(1-e^{\tfrac{\Xi_1}{r_{BL,T}}-\tfrac{\Xi_1}{r_p}}\right) \right..
\end{align}
Additionally, the total radial mass and heat transfer from the droplet to the far field are presented in Tab.~\ref{tab:flows}.

To evaluate the  transfer rates, one needs to know the mass and heat transfer film thickness. 
These thicknesses are commonly inferred from the empirical heat transfer correlations of spheres, which do not include Stefan flow. 
The correlations are formulated to find the Nusselt number $Nu_{m,0} = 2\left(1-\tfrac{r_p}{r_{BL,T}}\right)$ such, that: $\dot{Q}_r^{TTD} =  \pi \lambda_{m} d_p \left(T_p-T_s\right) Nu_{m,0}$.
Thus, the film thickness is: $\delta_T = d_p / \left(Nu_{m,0}-2\right)$.

Fr\"{o}ssling~\cite{frossling1938uber} introduced an empirical correlation to assess Nusselt numbers
of spheres in forced convection in the form:
\begin{align}
\label{eq:RanzMarshallT}
Nu_{m,0} = 2 + C Re_m^{1/2} Pr_m^{1/3},
\end{align}
with $C=0.552$, Ranz and Marshall~\cite{ranz1952evaporation} reported $C=0.6$ in their empirical study. 
Throughout this work $C=0.6$ is retained in accordance with the Ranz-Marshall model.
\mynormaltext{The heat and mass transfer film thicknesses are treated analogously}:
\begin{align}
    \delta_T =  \tfrac{1}{0.6} Pr_m^{-1/3} Re_m^{-1/2} d_p; \quad \delta_M =  \tfrac{1}{0.6} Sc_m^{-1/3} Re_m^{-1/2} d_p,
\end{align}
where $Pr_m = \tfrac{c_{p,m} \mu_m}{\lambda_m}$ is the Prandtl number in the heat transfer film, 
$Sc_m = \tfrac{\mu_m}{\rho_{m} \mathcal{D}_{m}}$ is the Schmidt number of the vapor in the mass transfer film,
and $Re_m = \tfrac{ \rho_{m} \left|\mathbf{u}_s\right| d_p}{\mu_{m}}$ is the Reynolds number of the moving droplet with the gas phase viscosity: $\mu_{m}$, and the slip velocity between the moving droplet and the gas phase: $\mathbf{u}_s$. 
\mynormaltext{The mass transfer film thickness corresponds} to a Sherwood number of:
\begin{align}
    Sh_{m,0} = \tfrac{2}{1-\tfrac{r_p}{r_{BL,M}}} = 2 + 0.6 Re_m^{1/2} Sc_m^{1/3}.
\end{align}

Finally, it is implicitly assumed that the gas phase material properties are constant across the film around the droplet \mynormaltext{in the models described in this section}. 
However these properties vary in function of the temperature and composition \mynormaltext{in reality}. 
\mynormaltext{To overcome this difficulty, the assumption of the existence of mean state is used, such}
that using the properties of this state results in minimal error in momentum, heat, and mass transfer. 
The subscript "$m$" represents this mean state. 

Yuen and Chen~\cite{yuen1976drag} propose the so called "1/3 law", where the mean properties are evaluated at a virtual state characterized by a weighted average of the seen and interface composition and temperature:
\begin{align}
\label{eq:OneThirdClassic}
T_m = \alpha T_s + (1-\alpha) T_p; \quad
Y_{k,m} = \alpha Y_{k,s} + (1-\alpha) Y_{k,i},
\end{align}
with $\alpha=1/3$, and $k=1..S$ where $S$ is the number of species considered in the gas phase. 
Recently this method is applied in most Lagrangian spray combustion simulations\mynormaltext{ \cite{jenny2012modeling,noh2018comparison,sitte2019large,benajes2021analysis}, and it is used throughout the present study}. 
To evaluate the transfer rates at a given far field and droplet surface conditions, the material properties used in the above equations \mynormaltext{($c_{p,vap,m}$, $c_{p,m}$, $\lambda_{m}$, $\mu_{m}$, $\rho_{m}$, $\mathcal{D}_{m}$)} need to be calculated according to the "1/3 law".

The specific heat of the vapor and of the mean gas mixture is calculated based on the NASA polynomials widely used in reacting flow calculations. 
\mynormaltext{The specific heat of the pure vapor is typically higher than that of the mixture} for the studied complex hydrocarbon fuels, meaning, that the factor ${c_{p,vap,m} }/{c_{p,m}}$ of $\Xi_1$ is above unity. 

The transport properties: thermal conductivity $\lambda_{m}$, dynamic viscosity $\mu_{m}$, and the diffusivity of the volatile species $\mathcal{D}_{m}$, are calculated following \mynormaltext{the transport theory of multicomponent mixtures.}
And the mixture averaged molar diffusivity of the volatile species $\mathcal{D}_m^{mol}$ is used to yield \mynormaltext{the \emph{mass} diffusivity} $\mathcal{D}_{m}$, according to Ebrahimian and Habchi~\cite{ebrahimian2011towards}:
\begin{align}
    \mathcal{D}_m = \mathcal{D}_m^{mol} \left(1-X_{f,m}+Y_{f,m} \sum_{k=1,k \ne f}^{S} \dfrac{X_{k,m}}{Y_{k,m}}\right),
\end{align}
where $X_{k,m}$ is the molar fraction of species $k$ in the mean mixture.

Finally, the density is evaluated using the ideal gas law:$\rho_{m} = \left(P W_m\right)/\left(R_u T_m\right)$,
where $P$ is the pressure of the system, $W_m$ is the mean molar mass, and $R_u$ is the universal gas constant.
\mynormaltext{The fuel's mass diffusivity and mean density are crucial properties for the evaporation process, as the mass flow rates are directly proportional to $\rho_{m}\mathcal{D}_m$, and this term is also present in the Lewis number, influencing the Stefan flow effects through the factor $\Xi_1$. 
The less volatile hydrocarbon fuels are generally also characterized by lower diffusivity, further impeding the evaporation.
}

\subsection{Droplet evaporation models}

In the present modeling framework, the droplets are treated as homogeneous spheres under the infinite conductivity assumption. 
I.e.: the heat (and mass) transfer inside the droplet is significantly faster than outside of it, thus the droplet can be characterized by a constant temperature profile. 
\mynormaltext{The Biot number $Bi$ provides a comparison of the time scales of heat transfer outside and inside the droplet: $Bi=\tfrac{h d_p}{\lambda_{p}}$, where the $h$ is the heat transfer coefficient in the gas phase: $h = \tfrac{-\dot{Q}_r}{\pi d_p^2 \left( T_s-T_p \right)} $ and $\lambda_{p}$ is the thermal conductivity in the liquid phase. Thus, the Biot number is $Bi=\tfrac{\lambda_m Nu_m}{\lambda_p}$, where $Nu_m$ is an effective Nusselt number, that may be corrected for considering Stefan flow effects. 
The liquid thermal conductivity is and order of magnitude higher, than the mean gas phase thermal conductivity, and $Nu_m$ is generally low if the effect of Stefan flow is considered, thus the Biot number is low, and the infinite conductivity model is valid.}
In this approach, the droplet is fully described by two quantities 
influencing the evaporation: its mass and its specific enthalpy or temperature.

Ordinary differential equations can be formed to represent the conservation of these quantities in relation to the transfer rates presented in subsection~\ref{subsec:quasi_steady_heat_and_mass}.
This model postulates, that while the inner droplet temperature profile relaxes to a constant temperature infinitely fast, similarly the gas phase temperature and vapor mass fraction profiles also relax to their steady state infinitely fast (quasi-steady assumption).
The mass change of the droplet is simply expressed as:
\begin{align}
    \label{eq:mass_ODE}
    \dfrac{\mathrm{d} m_p}{\mathrm{d}t} =  -\dot{m}_r,
\end{align}
where $m_p$ is the mass of the droplet.
\mynormaltext{
The energy conservation of a droplet can be formulated in terms of the droplet temperature as:
\begin{align}
    \label{eq:temperature_ODE}
    \dfrac{\mathrm{d}T_p}{\mathrm{d}t} = 
    \dfrac{-\dot{Q}_r}{m_p c_{p,p}} 
    + \dfrac{L_v}{m_p c_{p,p}} \dfrac{\mathrm{d} m_p}{\mathrm{d}t},
\end{align}
using the definition of the isobaric specific heat $c_{p,p} = \tfrac{\partial h_p}{\partial T}_{|p}$ and the latent heat of evaporation: $L_v = h_v-h_p$, with $h_p$ and $h_v$ being the liquid and vapor enthalpies respectively.  }

The remaining unknown terms of the heat and mass conservation equations 
are closed using the standardized material property functions of Daubert and Danner~\cite{daubert1985data}.
The necessary properties are the density of the droplet $\rho_p$ relating the droplet mass to the diameter, 
the specific heat of the liquid phase $c_{p,p}$, and the latent heat of evaporation $L_v$. 
Furthermore, in the 
\mynormaltext{evaporation models defined below}
, with the exception of \mynormaltext{the non-equilibrium models}, the interface vapor mass fraction is related to the droplet temperature assuming local thermodynamic equilibrium: $Y_{f,i} = Y_{f,i}^{eq}$. 
The saturation pressure $P_{sat}$ is also evaluated using the functions of Daubert and Danner~\cite{daubert1985data}.
In accordance with Rault's law, the equilibrium vapor mole fraction on the droplet interface is given by $X_{f,i}^{eq} = P_{sat}/P$. 
To yield the equilibrium interface vapor mass fraction, the frozen chemistry assumption is used, postulating that the bath gas composition is constant in the boundary layer around the droplet: 
$Y_{f,i}^{eq} =  X_{f,i}^{eq} / \left[ X_{f,i}^{eq} + \left(1-X_{f,i}^{eq}\right) \tfrac{W_b}{W_f} \right]$,
where $W_b$ is the mean molar mass of the bath gas, and $W_f$ is the molar mass of the volatile component. 
\mynormaltext{The frozen chemistry assumption  speculates, that the chemical reactions are inactive in the thin boundary layer surrounding the droplet, thus the bath gas species do not react with the volatile fuel in the film, and the conservation equations of fuel mass Eq.~\eqref{eq:radial_mass_transfer_rate} and enthalpy Eq.~\eqref{eq:radial_heat_transfer_rate} only need to consider advection and diffusion as derived in subsection~\ref{subsec:quasi_steady_heat_and_mass}. }

The different evaporation models used in this work are summarized in Tab.~\ref{tab:evaporation_models} and further described below. 
They differ in terms of considering Stefan flow, 
introducing additional corrections for the film thickness, 
and considering non-equilibrium conditions on the liquid-vapor interface.

\subsubsection{Diffusion only model (D/D: MTD + TTD)}

The Diffusion only model considers the diffusion based transport quantities derived in subsection~\ref{subsec:quasi_steady_heat_and_mass}.
The mass and heat transfer rates are given by $\dot{m}_r^{MTD}$ and $\dot{Q}_r^{TTD}$, thus both rates scale linearly with the "potential differences": $\left(Y_{f,i}-Y_{f,s}\right)$ and $\left(T_p-T_s\right)$.
This model is equivalent to M5 (Mass analogy IIa) of Miller~et~al.~\cite{miller1998evaluation}.

\subsubsection{Classical model (S/D: MTS + TTD)}

The Classical model combines the mass transport considering Stefan flow ($\dot{m}_r^{MTS}$) with the thermal transport neglecting Stefan flow ($\dot{Q}_r^{TTD}$). 
Such a combination is quite straightforward, as it is more natural to solve the mass transfer including Stefan flow (unimolecular diffusion), while in heat transfer the Stefan flow effects are not inherent to the problem. 
Nevertheless, this asymmetry makes the Classical model (S/D) open to doubt.
\mynormaltext{In this case the mass transfer rate is no longer proportional to the difference between fuel mass fractions on the interface and in the far-filed, but the rate is governed by the logarithmic term: $\ln\left(\tfrac{1-Y_{f,s}}{1-Y_{f,i}}\right)$. 
This term is widely expressed as $\ln\left( 1 + B_M\right)$, giving the definition of the Spalding mass transfer number: $B_M = \tfrac{Y_{f,i} - Y_{f,s}}{1-Y_{f,i}}$. 
The transformation is preferred, since $B_M$ expresses the mass transfer potential in a single variable. 
It tends to zero at low evaporation rates, and it provides a more sensitive measure near the boiling point of the droplet, since $B_M$ can reach very high values, as the interface vapor mass fraction approaches $1$.} 
This model is equivalent to M1 (Classical rapid mixing) of Miller~et~al.~\cite{miller1998evaluation}, and to EM1 of Noh~et~al.~\cite{noh2018comparison}.
\mynormaltext{The mass conservation equation of the classical model is shown in Tab.~\ref{tab:evaporation_models}.}

\subsubsection{Bird's correction (B: MTS + TTS)}

Bird~et~al.~\cite[§19.4,§22.8]{bird1960transport} noted, 
that high mass transfer rates distort the boundary layer profiles, 
as the energy carried by the \mynormaltext{ unimolecular diffusion of vapor} becomes significant.
In the case of forced convection, this results in decreased heat transfer rates if the net mass transfer is away from the surface (e.g.: fast evaporation of a droplet).

Bird~et~al. defined a rate factor $\beta$, as the ratio of 
enthalpy transported by Stefan flow 
to the enthalpy transported by conduction in the absence of the Stefan flow:
\begin{align}
\label{eq:BirdBeta}
\beta = \dfrac{c_{p,vap,m} \bm{\Phi}_{M,f,r}^{MTS} \left( T_p-T_s \right) }{ \bm{\Phi}_{T,r}^{TTD} },
\end{align}
By substituting the expressions of Tab.~\ref{tab:flows}, $\beta$ may be expressed as:
\begin{align}
\label{eq:AlternateBeta}
\beta = \dfrac{c_{p,vap,m} }{ c_{p,m} } \dfrac{Pr_m}{Sc_m} \dfrac{Sh_{m,0}}{Nu_{m,0}} \ln(1+B_M)
=\phi_m \ln(1+B_M),
\end{align}
where $\phi_m=\tfrac{c_{p,vap,m} }{ c_{p,m} } \tfrac{1}{Le_m} \tfrac{Sh_{m,0}}{Nu_{m,0}}$ expresses the $\beta$ ratio's dependence on factors other than the Spalding number, and $Le_m = \tfrac{Sc_m}{Pr_m}$ is the Lewis number of the vapor in the mixture based on the mass diffusivity $\mathcal{D}_m$.

Bird~et~al. propose an effective Nusselt number corrected for the effect of Stefan flow as:
\begin{align}
Nu_m^{*,B} = \dfrac{\beta}{e^{\beta}-1} Nu_{m,0}
=\dfrac{\phi_m \ln\left(1+B_M\right)}{\left(1+B_M\right)^{\phi_m}-1}Nu_{m,0},
\end{align}
based on the simultaneous heat and mass transfer of a flat plate using film theory.
Note, that $Nu_m^{*,B}$ does not have direct relation to the film thickness. 
It is merely a factor that represents the effect of film thickness and the effect of Stefan flow together.
Analogous to $B_M$, a Spalding \emph{heat} transfer number can be defined as:
$
1+B_T = \left(1+B_M\right)^{\phi_m}
$.
In case Bird's correction is used, the heat transfer equation of the droplet takes the form:
\begin{align}
\label{eq:HTBird}
\dfrac{\mathrm{d}T_p}{\mathrm{d}t} = 
\dfrac{\pi d_p \lambda_{g,m} Nu_{m,0}}{m_p c_{p,p}} \left(T_s-T_p\right) \dfrac{\ln\left(1+B_T\right)}{B_T}
+ \dfrac{L_v}{m_p c_{p,p}} \dfrac{\mathrm{d} m_p}{\mathrm{d}t}.
\end{align}
Bird~et~al.~\cite{bird1960transport} derived the above correction for mass transfer from a flat interface. 
However, one may \mynormaltext{demonstrate} that \mynormaltext{the same correction is to be applied} in spherical coordinates. 
\mynormaltext{Bird's correction considers the radial heat flux with Stefan flow $\dot{Q}_r^{TTS}$ defined in Eq.~\eqref{eq:TTS_heat_transfer_rate},}
the components of the last term of this equation are:
\begin{align}
2 \dfrac{\Xi_1}{r_p} 
= - \beta Nu_{m,0}; \quad
\dfrac{\Xi_1}{r_{BL,T}}-\dfrac{\Xi_1}{r_p} 
= \beta.
\end{align}
Hence, despite the different solution, the correction proposed by Bird
still holds for spheres as well: \mynormaltext{$
Nu_m^{*,B} / Nu_{m,0} = $ $2 \tfrac{\Xi_1}{r_p} / \left(1-e^{\tfrac{\Xi_1}{r_{BL,T}}-\tfrac{\Xi_1}{r_p}}\right)$  
$= \beta  / \left(e^{\beta} - 1 \right)$}.

\subsubsection{Abramzon-Sirignano model (AS)}

Abramzon and Sirignano~\cite{abramzon1989droplet} argue, that the \mynormaltext{heat and mass transfer} film thickness is influenced by Stefan flow. 
The difference between the affected and unaffected film thickness is expressed by the correction factors \mynormaltext{$F_T = \delta_T^*/\delta_T$ and $F_M = \delta_M^*/\delta_M$},
where the $*$ superscript signifies the film thickness in the presence of Stefan flow.
In their study of a vaporizing wedge, they concluded, that $F_T$ and $F_M$ are mainly influenced by the transfer numbers $B_M$ and $B_T^*$.
The correction factors take the form: 
\begin{align}
\label{eq:ASFactor}
F(B) = \left(1+B\right)^{0.7} \dfrac{\ln\left(1+B\right)}{B},
\end{align}
\mynormaltext{where $F(B)$ can be $F_T(B_T^*)$ or $F_M(B_M)$, with $B_T^*$ evaluated using the modified Sherwood and Nusselt numbers as detailed below.}
Note, that the validity range of Eq.~\eqref{eq:ASFactor} is in $0\le B \le 20$. 
Strong Stefan flow may thicken the boundary layers by as much as $28\%$.

The modified Nusselt and Sherwood numbers take the form:
\begin{align}
\label{eq:ASNuSh}
Nu_m^{*,AS} = 2 + \dfrac{Nu_{m,0}-2}{F_T}; \quad Sh_m^{*,AS} = 2 + \dfrac{Sh_{m,0}-2}{F_M},
\end{align}
Finally, the model is closed, by relating the Spalding transfer numbers of mass and energy through \mynormaltext{
$B_T^* = \left(1+B_M\right)^{\phi_m^*}-1$,
}
where $\phi_m^* = \tfrac{c_{p,vap,m} }{ c_{p,m} } \tfrac{1}{Le_m} \tfrac{Sh_m^{*,AS}}{Nu_m^{*,AS}}$ is the parameter introduced in Eq.~\eqref{eq:AlternateBeta}, but evaluated at the modified Nusselt and Sherwood numbers. Thus, the Abramzon-Sirignano model is implicit and has to be solved iteratively, considering \mynormaltext{$\phi_m^* = \phi_m \tfrac{Sh_m^{*,AS}}{Sh_{m,0}} \tfrac{Nu_{m,0}}{Nu_m^{*,AS}}$}.

Sazhin~\cite{sazhin2014droplets} points out, that the naming of modified Nusselt and Sherwood numbers is a possible source of confusion, as the work of Abramzon and Sirignano departs from a model, that already considers the Stefan flow effects in both heat and mass transfer.
The correction introduced is regarding the film thicknesses only.
Stefan flow effects should be considered in the heat transfer as in Eq.~\eqref{eq:HTBird}.
Thus, the equations solved using the Abramzon-Sirignano \mynormaltext{take the form presented in Tab.~\ref{tab:evaporation_models}.}

\begin{figure}[h!]
    \centering
	\includegraphics[width= 0.95 \textwidth]{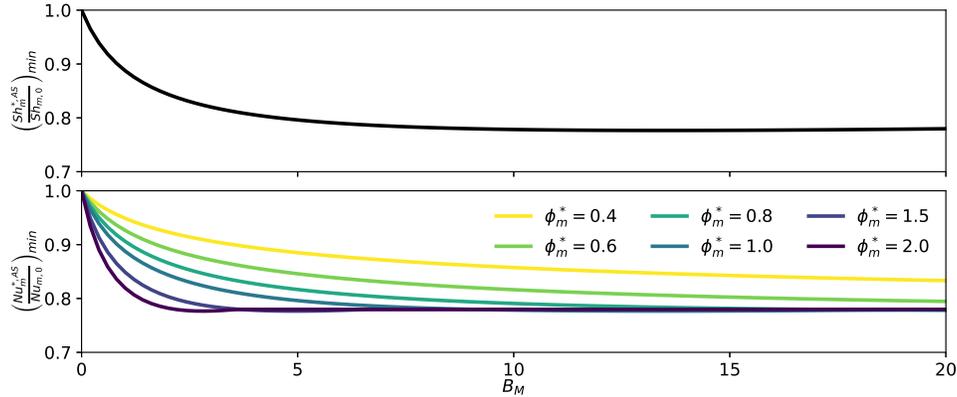}
    \caption{The minimum possible ratio of corrected and uncorrected Nusselt and Sherwood numbers of Abramzon and Sirignano~\cite{abramzon1989droplet} parametrized by $\phi_m^*=\tfrac{c_{p,vap,m} }{ c_{p,m} } \tfrac{1}{Le_m} \tfrac{Sh_m^{*,AS}}{Nu_m^{*,AS}}$.}
    \label{fig:AS_corr}
\end{figure}

Rearranging Eq.~\eqref{eq:ASNuSh} as:
\begin{align}
\dfrac{Nu_m^{*,AS}}{Nu_{m,0}} = \dfrac{1 + 2\tfrac{F_T-1}{Nu_{m,0}}}{F_T}; \quad
\dfrac{Sh_m^{*,AS}}{Sh_{m,0}} = \dfrac{1 + 2\tfrac{F_M-1}{Sh_{m,0}}}{F_M},
\end{align}
one can see, that in the limiting case of $Re_m \rightarrow 0$ \mynormaltext{($Nu_{m,0} \rightarrow 2$, $Sh_{m,0} \rightarrow 2$)}, there is no correction, irrespective of the transfer rates, since the "film" thickness at $Re_m \rightarrow 0$ approaches infinity. However, for high Nusselt and Sherwood numbers of fast moving droplets, the correction is limited by:
\mynormaltext{
$
\left(\tfrac{Nu_m^{*,AS}}{Nu_{m,0}}\right)_{min} = \tfrac{1}{F_T}
$,
$
\left(\tfrac{Sh_m^{*,AS}}{Sh_{m,0}}\right)_{min} = \tfrac{1}{F_M}.
$
}
The typical values of the correction at the high Reynolds number limit are illustrated in Fig.~\ref{fig:AS_corr}\mynormaltext{, showing that the Abramzon-Sirignano model can result in a maximum of $22\%$ further reduction of Nusselt and Sherwood numbers compared to Bird's correction.}
Note, that here $B_T$ is clipped at $20$ according to the validity range of Eq.~\eqref{eq:ASFactor}, affecting the curves of $\phi_m^* > 1$.

\subsubsection{Langmuir-Knudsen models (LK1,LK2)}

In the discussion above, the surface composition is determined 
using the equilibrium vapor pressure on the droplet surface, 
and the frozen chemistry assumption.
\mynormaltext{Former means, that the partial pressure of the volatile component on the droplet interface is the saturation pressure corresponding to the interface temperature. 
While latter refers to the assumed inactivity of chemical reactions in the mass transfer film.}
The Langmuir-Knudsen model considers an additional resistance in the mass transfer, 
by postulating that the vapor mole fraction on the droplet interface 
is not at equilibrium. 

The non-equilibrium mole fraction on the droplet surface may be calculated as:
\mynormaltext{
$
X_{f,i}^{neq} = X_{f,i}^{eq} - \tfrac{2 L_K}{d_p} \beta,
$
}
where $\beta$ can be evaluated:
\mynormaltext{
in model LK1 from the equilibrium mass transfer rate: $\beta^{eq} = \phi_m \ln\left(1+B_M^{eq}\right)$, 
or in model LK2 iteratively from the non-equilibrium mass transfer rate: $\beta^{neq} = \phi_m \ln\left(1+B_M^{neq}\right)$,
}
where $B_M^{eq}$ and $B_M^{neq}$ are the Spalding mass transfer numbers evaluated using the equilibrium and non-equilibrium surface mass fractions respectively, and
\mynormaltext{
$
L_K = \rho_{m} \mathcal{D}_{m} \sqrt{ 2 \pi T_p \frac{R_u}{W_f}} / \left(\alpha_e P\right)
$
is the Knudsen layer thickness, 
}
with $\alpha_e = 1$ molecular accommodation coefficient.

The non-equilibrium interface vapor mass fraction $Y_{f,i}^{neq}$ is still calculated with the frozen chemistry assumption, but replacing  $X_{f,i}^{eq}$ with $X_{f,i}^{neq}$.
In case the Langmuir-Knudsen model is used, the Spalding number is evaluated with the non-equilibrium interface vapor mass fraction. 
However, the representative gas phase properties are calculated assuming equilibrium conditions, thus an  \mynormaltext{additional} iterative lookup of the mean properties can be avoided. \cite{miller1998evaluation}
However, the calculation of the surface mole fraction requires an iterative solution for model LK2. 
\mynormaltext{
The models include Bird's correction, but using the non-equilibrium transfer numbers evaluated using the non-equilibrium surface mass fraction: $B_M^{neq} = \tfrac{Y_{f,i}^{neq} - Y_{f,s}}{1-Y_{f,i}^{neq}}$, and $B_T^{neq} = \left(1+B_M^{neq}\right)^{\phi_m}-1$.
}

\begin{table}[h!]
    \centering
    \begin{tabular}{r|c}
    D/D & 
    $\dfrac{\mathrm{d}T_p}{\mathrm{d}t} = 
\dfrac{\pi d_p \lambda_{m} Nu_{m,0}}{m_p c_{p,p}} \left(T_s-T_p\right) 
+ \dfrac{L_v}{m_p c_{p,p}} \dfrac{\mathrm{d} m_p}{\mathrm{d}t}$ \\
    &
    $\dfrac{\mathrm{d} m_p}{\mathrm{d}t} = -\pi d_p \rho_{m} \mathcal{D}_{m} Sh_{m,0} \left(Y_{f,i} - Y_{f,s}\right)$\\\hline
    S/D &
    $\dfrac{\mathrm{d}T_p}{\mathrm{d}t} = 
\dfrac{\pi d_p \lambda_{m} Nu_{m,0}}{m_p c_{p,p}} \left(T_s-T_p\right) 
+ \dfrac{L_v}{m_p c_{p,p}} \dfrac{\mathrm{d} m_p}{\mathrm{d}t}$ \\
    &
    $\dfrac{\mathrm{d} m_p}{\mathrm{d}t} = -\pi d_p \rho_{m} \mathcal{D}_{m} Sh_{m,0} \ln\left(1+B_M\right)$\\\hline
    B &
    $\dfrac{\mathrm{d}T_p}{\mathrm{d}t} = 
\dfrac{\pi d_p \lambda_{m} Nu_{m,0}}{m_p c_{p,p}} \left(T_s-T_p\right) \dfrac{\ln\left(1+B_T\right)}{B_T}
+ \dfrac{L_v}{m_p c_{p,p}} \dfrac{\mathrm{d} m_p}{\mathrm{d}t}$ \\
    &
    $\dfrac{\mathrm{d} m_p}{\mathrm{d}t} = -\pi d_p \rho_{m} \mathcal{D}_{m} Sh_{m,0} \ln\left(1+B_M\right)$\\\hline
    AS &
    $\dfrac{\mathrm{d}T_p}{\mathrm{d}t} = 
\dfrac{\pi d_p \lambda_{m} Nu_m^{*,AS}}{m_p c_{p,p}} \left(T_s-T_p\right) \dfrac{\ln\left(1+B_T\right)}{B_T}
+ \dfrac{L_v}{m_p c_{p,p}} \dfrac{\mathrm{d} m_p}{\mathrm{d}t}$ \\
    &
    $\dfrac{\mathrm{d} m_p}{\mathrm{d}t} = -\pi d_p \rho_{m} \mathcal{D}_{m} Sh_m^{*,AS} \ln\left(1+B_M\right)$\\\hline
    LK &
    $\dfrac{\mathrm{d}T_p}{\mathrm{d}t} = 
    \dfrac{\pi d_p \lambda_{m} Nu_{m,0}}{m_p c_{p,p}} \left(T_s-T_p\right) \dfrac{\ln\left(1+B_T^{neq}\right)}{B_T^{neq}}
    + \dfrac{L_v}{m_p c_{p,p}} \dfrac{\mathrm{d} m_p}{\mathrm{d}t}$ \\
    &
    $\dfrac{\mathrm{d} m_p}{\mathrm{d}t} = -\pi d_p \rho_{m} \mathcal{D}_{m} Sh_{m,0} \ln\left(1+B_M^{neq}\right)$
    \end{tabular}
    \caption{Summary of the different evaporation models. D/D: diffusion only model, S/D: Classical model, B: Bird's correction, AS: Abramzon-Sirignano model, LK: Langmuir-Knudsen model.}
    \normalsize
    \label{tab:evaporation_models}
\end{table}

\section{Single droplet behavior}

\subsection{Wet-bulb conditions}
\label{subsec:wet_bulb}

In psychrometry, the \emph{thermodynamic} web-bulb temperature is defined as the temperature of adiabatic saturation, 
i.e.: the temperature to which a given fuel/bath gas mixture can be adiabatically cooled 
by the evaporation of the fuel at the same temperature into the vapor/bath gas mixture~\cite{gatley2005understanding}.

\begin{figure}[h!]
    \centering
    \centering
	\begin{tikzpicture}
	\draw[thick,-] (0,1) -- (3,1);
	\draw[thick,-] (3,1) -- (3,0);
	\draw[thick,-] (9,0) -- (9,1);
	\draw[thick,-] (9,1) -- (14,1);
	\fill[pattern=north west lines] (0,0.5) rectangle (3,1);
	\fill[pattern=north west lines] (2.5,0.5) rectangle (3,0);
	\fill[pattern=north west lines] (9,0) rectangle (9.5,1);
	\fill[pattern=north west lines] (14,0.5) rectangle (9.5,1);
	\draw[thick,-] (0,3) -- (14,3);
	\fill[pattern=north west lines] (0,3) rectangle (14,3.5);
	\shade[top color=blue!30,bottom color=white] (3.02,0.02) rectangle (8.98,0.8);
	\draw[very thick,-latex] (1,1.5) -- (2.5,1.5) node[midway,above] {$\dot{m}$};
	\node[above]  at (6,1.5){$\dot{m}_f$};
	\draw[very thick,-latex] (10,1.5) -- (11.5,1.5) node[midway,above] {$\dot{m}+\dot{m}_f$};

	\foreach \x [evaluate=\x as \y using 1.8-(\x-3)*0.155] in {3,3.5,...,8.5}
		\draw[-latex] (\x,0.8)   -- (\x,  \y);

	\node[above]  at (1.75,2) {$Y_{f,s},T_s$};
	\node[above]  at (10.75,2){$Y_{f,i}^{th},T_p^{th}$};

	\node  at (6,0.4){$T_p^{th}$};

	\end{tikzpicture}
    \caption{Illustration of the thermodynamic wet-bulb temperature definition.}
    \label{fig:therm_wet_bulb_illustration}
\end{figure}
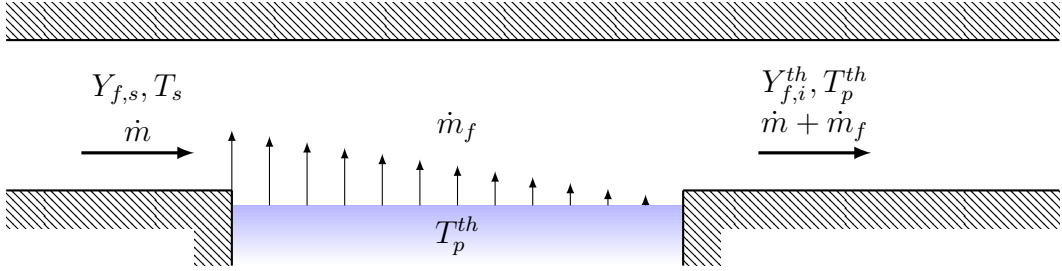

Figure~\ref{fig:therm_wet_bulb_illustration} illustrates the concept of adiabatic saturation using the nomenclature of this work, 
where $T_p^{th}$ is the thermodynamic wet-bulb temperature, $Y_{f,s}$ and $T_s$ are the fuel mass fraction and temperature at the studied gas conditions, while $Y_{f,i}^{th}$ is the fuel mass fraction at saturated conditions.
At the outlet of the control volume, the gas phase flow is in equilibrium with the liquid reservoir: \mynormaltext{i)} the liquid and the gas phase are at the same temperature $T_p^{th}$,
\mynormaltext{ii)} the partial pressure of fuel in the gas is the saturation pressure at $T_p^{th}$.
The model problem is characterized by an inlet mass flow rate of $\dot{m}$.
The mass flow rate of evaporation $\dot{m}_f$ is such, that $Y_{f,i}$ is reached at the outlet:
\begin{align}
\dot{m}_f &= \dfrac{Y_{f,i}^{th}-Y_{f,s}}{1-Y_{f,i}^{th}} \dot{m}.
\end{align}
The heat transfer to the liquid reservoir solely facilitates the evaporation, 
thus the energy conservation takes the form:
\begin{align}
\dot{m} \left(h_s\left(T_s\right)-h_s\left(T_p^{th}\right)\right) &= \dot{m}_f L_v, \\
\label{eq:ThermoWetBulb}
h_s\left(T_s\right)-h_s\left(T_p^{th}\right) &= B_M^{th}  L_v ,
\end{align}
where $h_s(T_s)$ is the enthalpy at the inlet, and $h_s(T_p^{th})$ is the enthalpy at the inlet composition but evaluated at the thermodynamic wet-bulb temperature.
Eq.~\eqref{eq:ThermoWetBulb} may be solved for $T_p^{th}$ at the given inlet conditions.

\begin{figure}[h!]
    \centering
	\includegraphics[width= 0.95 \textwidth]{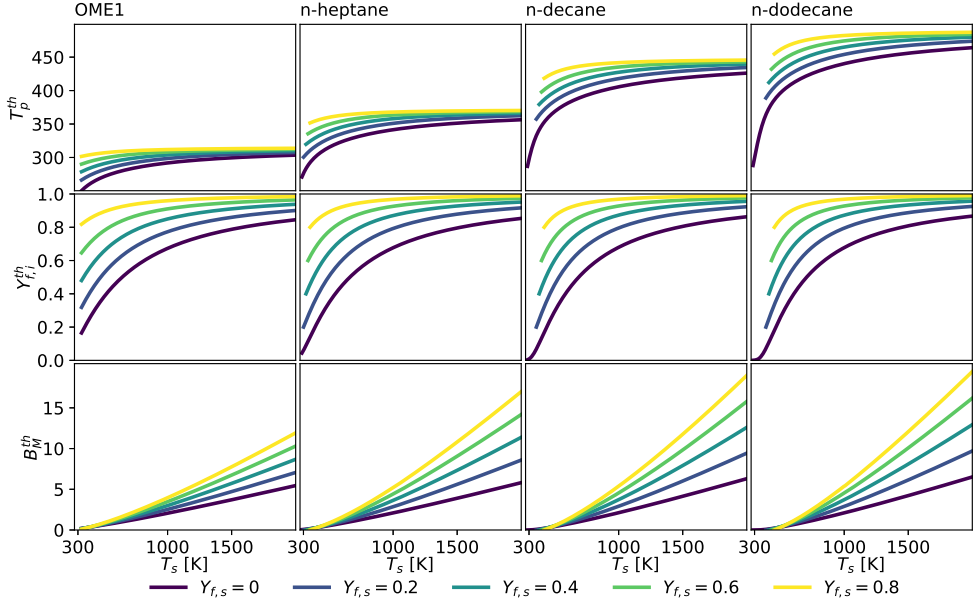}
    \caption{Thermodynamic wet-bulb conditions of OME1, n-heptane, n-decane, and n-dodecane at atmospheric pressure with air as bath gas according to Eq.~\eqref{eq:ThermoWetBulb}.}
    \label{fig:therm_wet_bulb}
\end{figure}

Figure~\ref{fig:therm_wet_bulb} shows the solutions of Eq.~\eqref{eq:ThermoWetBulb} at atmospheric pressure for OME1, n-heptane, n-decane, and n-dodecane.
The wet-bulb temperature $T_p^{th}$, the corresponding vapor mass fraction $Y_{f,i}^{th}$, and the Spalding mass transfer number $B_M^{th}$ are presented 
as function of the inlet temperature $T_s$, and the inlet vapor mass fraction $Y_{f,s}$.

The wet-bulb temperature asymptotically approaches the boiling point of the fluid as the inlet temperature and vapor mass fraction increase.
Consequently, the wet-bulb vapor mass fraction approaches unity. 
\mynormaltext{The conditions are evaluated under seen gas temperatures ranging from $300 \ \mathrm{K}$ to $2000 \ \mathrm{K}$. 
Concentrating on the $Y_{f,s}=0$ cases, one can identify that the volatility of the different fuels has the greatest effect at low seen gas temperatures. 
While in case of OME1, the wet-bulb vapor mass fraction $Y_{f,i}^{th}$ is already $\sim 0.18$ at $T_s=300 \ \mathrm{K}$, it is closer to zero in case of the other fuels, and n-decane and n-dodecane even show an inflection point in the wet-bulb vapor mass fraction.}

The \mynormaltext{proposed} experiment of Fig.~\ref{fig:therm_wet_bulb_illustration} is \mynormaltext{defined with the following hypotheses: the domain} is adiabatic to the environment, the liquid surface is large enough to reach equilibrium at the outlet, and the liquid is at a constant temperature equal to the outlet temperature. 
A more practical point of view is given by 
the \emph{psychrometric} wet-bulb temperature, that is defined by finding the equilibrium solution of the droplet 
at $\tfrac{\mathrm{d}T_p}{\mathrm{d}t}=0$, where the received heat is exactly the heat necessary for the phase transition. 
The equations yielding the wet-bulb conditions for the different models are described below and summarized in Tab.~\ref{tab:wet_bulb_for_models}.

\begin{table}[h!]
    \centering
    \begin{tabular}{r|c}
    Thermodynamic & 
    $h_s\left(T_s\right)-h_s\left(T_p^{th}\right) = B_M^{th}  L_v$\\
    Diffusion olny (D/D) & 
    $c_{p,vap,m} \left(T_s-T_p^{psy,D/D}\right) =  \phi_m \left( Y_{f,i}-Y_{f,s} \right) L_v$\\
    Classical (S/D) &
    $c_{p,vap,m} \left(T_s-T_p^{psy,S/D}\right) =  \ln\left(1+B_T^{psy,S/D}\right) L_v$\\
    Bird's correction (B) &
    $c_{p,vap,m} \left(T_s-T_p^{psy,B}\right) = B_T^{psy,B} L_v$\\
    Abramzon-Sirignano (AS) &
    $c_{p,vap,m} \left(T_s-T_p^{psy,AS}\right) = \dfrac{\phi^*_m}{\phi_m} 
    B_T^{psy,AS} L_v$\\
    Langmuir-Knudsen (LK) &
    $c_{p,vap,m} \left(T_s-T_p^{psy,LK}\right) = B_T^{neq,psy,LK} L_v$
    \end{tabular}
    \caption{Summary of the wet-bulb conditions of different evaporation models. D/D: diffusion only model, S/D: Classical model, B: Bird's correction, AS: Abramzon-Sirignano model, LK: Langmuir-Knudsen model.}
    \label{tab:wet_bulb_for_models}
    \normalsize
\end{table}

\mynormaltext{
The diffusion only model (D/D) is only able to produce equilibrium conditions for a limited range of seen temperatures. 
The inadequacy of the model is demonstrated in \ref{sec:wet_bulb_DD}.
}
The application of \mynormaltext{the diffusion only model} should be limited to low temperature, however choosing it over the other models presented here \mynormaltext{cannot be justified}.
\mynormaltext{
The classical evaporation model (S/D) is also flawed due to the arbitrary consideration of Stefan flow in only the mass transfer.
It is able to produce steady wet-bulb states at any seen temperature, as explained in \ref{sec:wet_bulb_SD}, 
however the resulting equilibrium conditions are unrealistic.  
}
This inconsistency is masked by the low significance of Stefan flow at low temperature applications where $\left(B_M \approx \ln\left(1+B_M\right) \right)$, however the model is often extended to regimes where Stefan flow dominates the overall heat transfer, resulting in highly overestimated evaporation rates.
\mynormaltext{
These models are given less attention in the rest of the present study, as their inherent flaws are are already demonstrated. 
}

The heat and mass transfer corrections of Bird~\cite{bird1960transport}, Abramzon and Sirignano~\cite{abramzon1989droplet}, and the Langmuir-Knudsen model are considered below. 
Based on the $\tfrac{\mathrm{d}T_p}{\mathrm{d}t}=0$ condition, the wet-bulb conditions of Bird's correction are simply given by:
\begin{align}
\label{eq:PsychroWetBulbBird}
c_{p,vap,m} \left(T_s-T_p^{psy,B}\right) &= B_T^{psy,B} L_v,
\end{align}
\mynormaltext{where $B_T = \left(1+B_M\right)^{\phi_m}-1$ is the Spalding heat transfer number.}
Meanwhile, the wet-bulb conditions for the model of Abramzon and Sirignano 
are determined by the equation:
\begin{align}
c_{p,m} \left(T_s-T_p^{psy,AS}\right) &= \dfrac{1}{Le_m} \dfrac{Sh_m^{*,AS}}{Nu_m^{*,AS}} 
\dfrac{B_T^{psy,AS}}{\ln\left(1+B_T^{psy,AS}\right)} \ln\left(1+B_M^{psy,AS}\right) L_v,
\end{align}
that can be expressed as:
\begin{align}
\label{eq:SprayModelPsychroWetBulbAS}
c_{p,m} \left(T_s-T_p^{psy,AS}\right) &= \dfrac{\phi^*_m}{\phi_m} 
    B_T^{psy,AS} L_v,
\end{align}
using the definitions of \mynormaltext{$\phi^*_m= \tfrac{c_{p,vap,m} }{ c_{p,m} } \tfrac{1}{Le_m} \tfrac{Sh_m^{*,AS}}{Nu_m^{*,AS}}= \phi_m \tfrac{Sh_m^{*,AS}}{Sh_{m,0}} \tfrac{Nu_{m,0}}{Nu_m^{*,AS}}$}. 

The wet-bulb conditions of Bird's correction and the Abramzon-Sirignano model are almost identical in the studied cases, since the ratio of corrected and uncorrected Sherwood and Nusselt numbers are rather similar for the two approaches. 
Note, that this similarity only concerns the wet-bulb conditions, the two models do differ in heat and mass transfer rate for non-zero Reynolds numbers.
For the Langmuir-Knudsen model (that includes Bird's correction), the wet-bulb conditions can be defined as: 
\begin{align}
\label{eq:PsychroWetBulbLK}
c_{p,vap,m} \left(T_s-T_p^{psy,LK}\right) &= B_T^{neq,psy,LK} L_v,
\end{align}
considering that the Spalding heat transfer number is based on the non-equilibrium vapor mass fractions.

\begin{figure}[h!]
    \centering
	\includegraphics[width= 0.95 \textwidth]{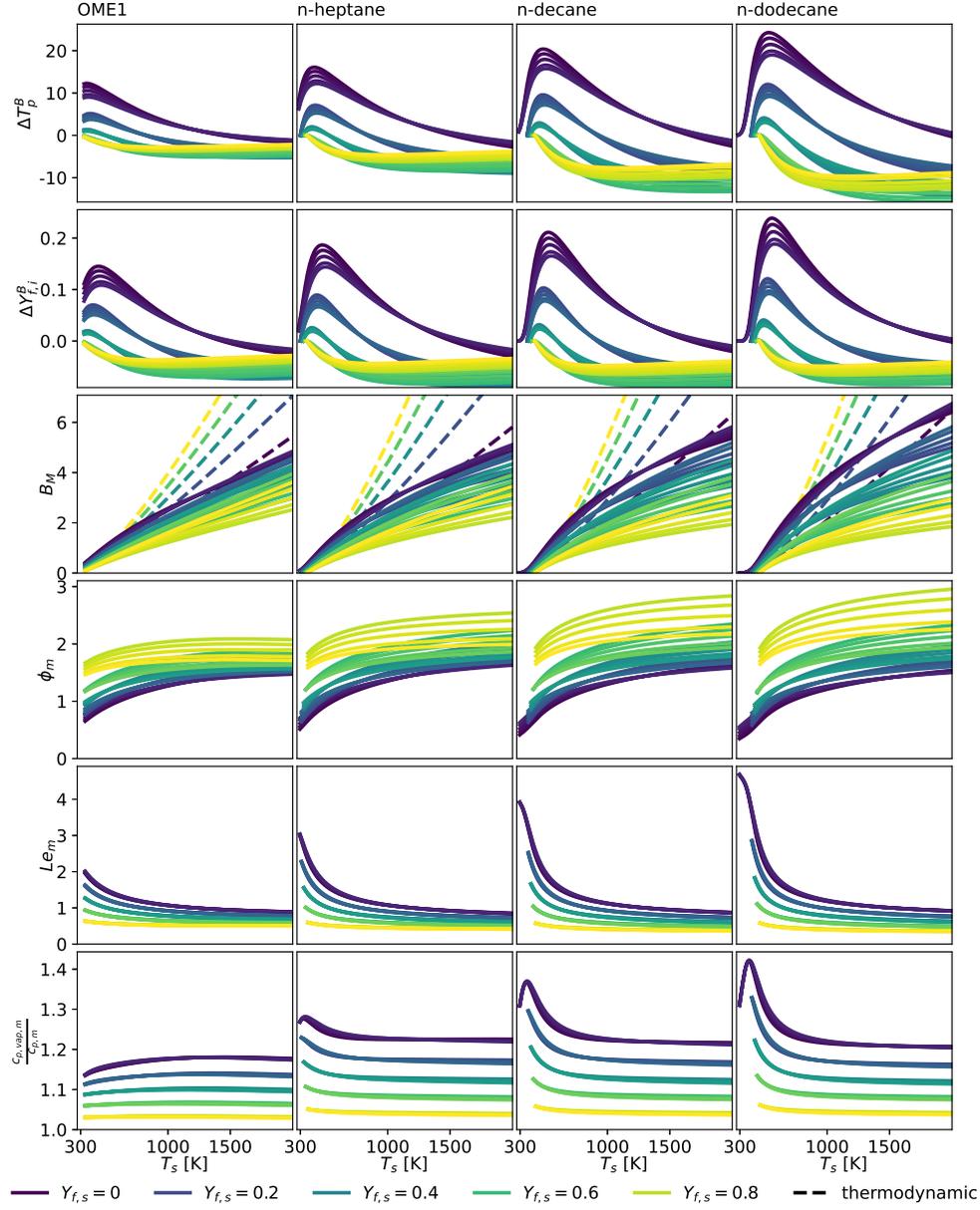}
    \caption{Comparison of psychrometric and thermodynamic wet-bulb conditions of OME1, n-heptane, n-decane, and n-dodecane at atmospheric pressure with air as bath gas considering Bird's correction according to Eq.~\eqref{eq:PsychroWetBulbBird}. The difference in wet-bulb temperatures and the corresponding vapor mass fractions are $\Delta T_p^{B} = T_p^{psy,B} - T_p^{th}$, and $\Delta Y_{f,i}^{B} = Y_{f,i}^{psy,B} - Y_{f,i}^{th}$.}
    \label{fig:psych_wet_bulb_bird}
\end{figure}

For simplicity, the results of the Abramzon-Sirignano model are not shown, since these are virtually the same as the results of Bird's correction displayed in
Fig.~\ref{fig:psych_wet_bulb_bird}.  
\mynormaltext{Likewise, the Langmuir-Knudsen model also produces similar equilibrium conditions to Bird's correction in case of large droplet diameters. The influence of the droplet diameter on this model is further discussed in \ref{sec:limit_LK}, showing that equilibrium conditions do not exist below a certain diameter.} 
\mynormaltext{
The wet-bulb conditions of Eq.~\eqref{eq:PsychroWetBulbBird} are compared to the thermodynamic wet-bulb state through
}
the difference in wet-bulb temperature $\Delta T_p^B=T_p^{psy,B}-T_p^{th}$ and vapor mass fraction $\Delta Y_{f,i}^B=Y_{f,i}^{psy,B}-Y_{f,i}^{th}$.

The quantities: $\Delta T_p^B$, $\Delta Y_{f,i}^B$, and $B_M$ of  Fig.~\ref{fig:psych_wet_bulb_bird} illustrate how the 
\mynormaltext{similarity} 
between heat and mass transfer equations is restored by considering the effect of Stefan flow on both transfer rates. 
Thus, the wet-bulb conditions generally get closer to the thermodynamic ones. 
The remaining differences 
\mynormaltext{between Eq.~\eqref{eq:ThermoWetBulb} and Eq.~\eqref{eq:PsychroWetBulbBird}} 
are mainly caused by the effect of Lewis number, and the disparity between the mean gas specific heat and the vapor specific heat both displayed in Fig.~\ref{fig:psych_wet_bulb_bird}. 

The parameter \mynormaltext{$\phi_m=\tfrac{c_{p,vap,m} }{ c_{p,m} } \tfrac{1}{Le_m} \tfrac{Sh_{m,0}}{Nu_{m,0}}$} combines \mynormaltext{these} Lewis number \mynormaltext{$\left(Le_m = \tfrac{\lambda_m}{c_{p,m} \rho_m \mathcal{D}_m}\right)$} and specific heat  effects.
\mynormaltext{Former parameter expresses the diffusivity of the volatile species relative to the thermal diffusivity, i.e.: high Lewis numbers correspond to low fuel diffusivity. 
While latter provides a measure of the heat carried by the unimolecular diffusion of the fuel (Stefan flow) compared to heat carried by other advective phenomena where all species are carried by the flow equally.}
\mynormaltext{However, the parameter $\phi_m$} depends also on the Reynolds number, as $Nu_{m,0}$ and $Sh_{m,0}$ are present in this parameter. 
Using the Fr\"{o}ssling-type correlations for $Nu_{m,0}$ and $Sh_{m,0}$ such as the Ranz-Marshall model, $\phi_m$ is limited between the  $\tfrac{c_{p,vap,m} }{ c_{p,m} } \tfrac{1}{Le_m}$ and $\tfrac{c_{p,vap,m} }{ c_{p,m} } \tfrac{1}{Le_m^{2/3}}$ corresponding to the low and high Reynolds number limits respectively. 
As Fig.~\ref{fig:psych_wet_bulb_bird} shows, the mass-based Lewis number of the volatile component drops sharply as the seen vapor mass fraction and temperature increase, and the specific heat ratio $\tfrac{c_{p,vap,m} }{ c_{p,m} }$ also shows more variation at low seen temperatures, while it is almost constant otherwise.
These two distinct regions are the most pronounced in case of the heavier hydrocarbons\mynormaltext{, that are characterized by near-zero wet-bulb vapor mass fractions at low seen temperatures}. 
The change of behavior with increasing seen temperature is explained by the changes in mean composition, 
since for high seen temperatures the mean composition $Y_{k,m}$ is practically constant because the interface composition approaches pure volatile vapor, while the mean temperature keeps increasing according to the "1/3 law". 
The mass-based Lewis number drops sharply as the mass fraction of vapor increases in the mean gas mixture, since it is largest in the dilute limit. 
Overall the high temperature region is dominated by high specific heat ratios and low Lewis numbers resulting in $\phi_m$ above unity. 
\mynormaltext{ If the far field does not contain any of the volatile species ($Y_{f,s}=0$), the Lewis numbers and specific heat ratios show a certain similarity across the different fuels at high seen temperatures. 
Thus under these conditions $\phi_m \approx 1.5$ is generally true for all fuels.  
However, as seen vapor mass fractions increases, ${c_{p,vap,m} }/{c_{p,m}}$ approaches unity slowly, while $Le_m$ decreases sharply, resulting in significantly higher $\phi_m$ for the heavier hydrocarbons.}

\subsection{Single droplet evaporation}
\label{subsec:single_drop}

As described above, most of the presented models can yield a psycrometric wet-bulb temperature, 
meaning that this temperature behaves as an attractor of the dynamic system formed by Eq.~\eqref{eq:mass_ODE} and Eq.~\eqref{eq:temperature_ODE}.
The diffusion only model (D/D) is limited in this sense, because it clearly does not have an equilibrium state for high seen temperatures, and the model is simply invalid for these cases.
The Langmuir-Knudsen models (\mynormaltext{LK1,}LK2) show signs of a similar problem, but only affecting very small droplets sizes.

Meanwhile, under the studied conditions, the mass of the droplet always approaches zero until the droplet completely evaporates. 
The mass conservation equation Eq.~\eqref{eq:mass_ODE} can be rewritten in terms of the diameter as:
\begin{align}
\dfrac{\mathrm{d} d_p^2 }{\mathrm{d}t}
&= 
- \dfrac{4 \dot{m}_r}{\pi \rho_p d_p}
- \dfrac{2 d_p^2}{3 \rho_p}  \dfrac{\mathrm{d} \rho_p  }{\mathrm{d}t}.
\end{align}
Since $\dot{m}_r$ scales linearly with the diameter as summarized in Tab.~\ref{tab:flows}, the droplet surface ($\sim d_p^2$) decreases at a constant rate, if the droplet temperature, Reynolds number, and the seen conditions are constant.
The evaporation of droplets that reached their equilibrium temperature are widely described using such "$d^2$" relations \cite{godsave1953studies}, simply implying that the evolution of droplet surface is linear in time:
\mynormaltext{
$
\tfrac{\mathrm{d} d_p^2 }{\mathrm{d}t} = -K,
$
}
where $K=-\left(\tfrac{\mathrm{d} m_p}{\mathrm{d}t}\right)^{psy} \tfrac{\pi \rho_p d_p}{4}$ is the vaporization rate constant. 
A droplet evaporation time can be defined as:
\mynormaltext{$\tau_{p,evap} = \tfrac{ d_{p,0}^2}{ K }$},
where $d_{p,0}$ is the initial droplet diameter, and $K$ is evaluated under the wet-bulb conditions.

\begin{figure}[h!]
    \centering
	
	\resizebox{0.95\textwidth}{!}{
	\begin{tikzpicture}
	           
	   \node[above right] at (0,0) {
	       \includegraphics[width= 10 cm]{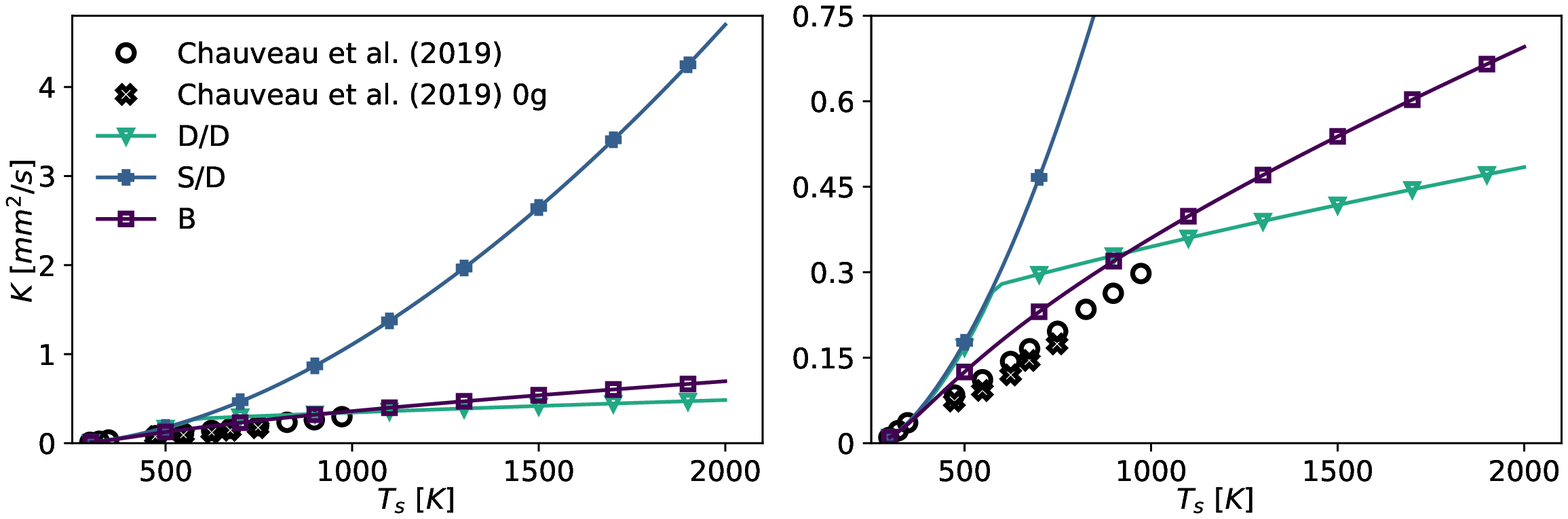}
	   };
	   
	   \node[above right] at (7.95,0.55) {
	       \includegraphics[width= 1.97 cm]{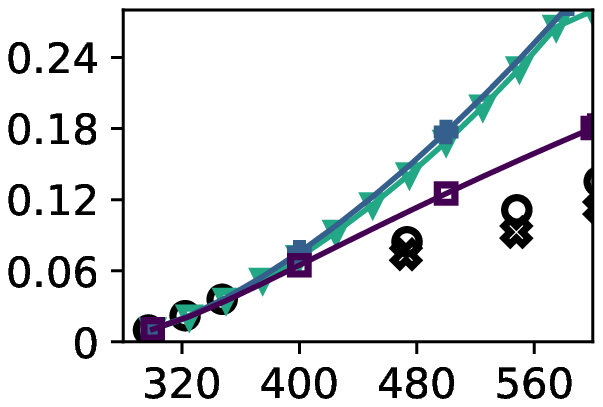}
	   };
	    
	\end{tikzpicture}
	}
	
    \caption{
    Comparison of the experimental evaporation rate constants of Chauveau~et~al.~\cite{chauveau2019analysis} against the diffusion only model (D/D), the classical model (S/D), and Bird's correction (B) for an n-heptane droplet of $d_{p,0} = 500 \ \mathrm{\mu m}$ in nitrogen gas atmosphere. The different figures show the data at different scales. The models are evaluated between $300 \ \mathrm{K}$ and $2000 \ \mathrm{K}$, with a step size of $25 \ \mathrm{K}$.
    }
    \label{fig:validation}
\end{figure}

Chauveau~et~al.~\cite{chauveau2019analysis} show that most vaporization rate constant measurements significantly overestimate $K$ at high temperature conditions, 
as additional heat is transferred to the droplet through support fibers that have a diameter comparable to the droplet diameter. 
They propose a measurement technique of reducing the support fiber diameter by an order of magnitude eliminating this deterministic measurement error.

Figure~\ref{fig:validation} shows the comparison of rate constants obtained for a stationary n-heptane droplet in nitrogen gas atmosphere under psychrometric wet-bulb conditions along different seen  gas temperatures for the diffusion only (D/D) and classical (S/D) models and for Bird's correction (B) along with the measurement data of Chauveau~et~al.~\cite{chauveau2019analysis}. 
The other models (AS,LK1,LK2) are omitted, since they are the same as Bird's correction (B) for large stationary droplets. 
In general, all models overestimate the experimentally determined evaporation rates at high temperature conditions. 

The best agreement with the measurement is observed using Bird's correction (B), that qualitatively captures the slope of $K$ as a function of the seen temperature. The remaining error is limited to a $20\%$ overestimation and can be attributed to the real gas behavior of the fluid in the heat and mass transfer films, as suggested by Ebrahimian and Habchi~\cite{ebrahimian2011towards}. \mynormaltext{However, addressing these effects} is out of the scope of the present study. 
The classical model (S/D) results in particularly fast evaporation, 
overestimating the evaporation rate by a factor of 3.5 for the highest temperature measurement ($T_s=973.15 \ \mathrm{K}$) 
and producing a 6.8 higher rate than Bird's correction (B) at $T_s=2000 \ \mathrm{K}$.
The diffusion only model (D/D) is assessed both in its range of applicability and outside of it. 
In the former regime, it follows closely the behavior of the classical model (S/D) despite the higher wet-bulb temperatures observed using the diffusion only model (D/D). 
The behavior of the model changes drastically, once thermal equilibrium conditions become impossible and $Y_{f,i}$ is clipped to 1. In this regime, the evaporation rates continue growing solely because $\rho_m\mathcal{D}_m$ increases due to the application of the "1/3-law". 
Note that this clipping is nonphysical, and simulations applying this model would violate energy conservation, as a significant part of the heat transferred to the droplet is not spent neither on increasing the droplet temperature, nor on facilitating the phase change.

\mynormaltext{
Comparing the vaportization rate constants $K=\mathcal{O}\left(0.1 \ \mathrm{mm^2/s}\right)$ of Fig.~\ref{fig:validation}, and the thermal diffusivity $\mathcal{D}_{t,m}=\mathcal{O}\left(10 \ \mathrm{mm^2/s}\right)$, one can see, that the quasi-steady state assumption of the heat and mass transfer processes is well-founded, since a unity Fourier number state $Fo=\tfrac{t \mathcal{D}_{t,m} }{d_p^2}$ is reached two orders of magnitude faster, than the time scale of evaporation. }

Depr{\'e}durand~et~al.~\cite{depredurand2010heat} and Castanet~et~al.~\cite{castanet2016evaporation} showed experimentally, 
that in case the initial droplet temperature is significantly lower than the wet-bulb temperature, then the majority of heat is transferred to the liquid phase and only a fraction of it facilitates the phase change.  
A scale of the heat-up time may be calculated using the initial heat-up rate:
\begin{align}
\tau_{p,heat} = \dfrac{T_{p}^{psy}-T_{p,0}}{\left(\tfrac{\mathrm{d}T_p}{\mathrm{d}t}\right)_{0}} 
=
\dfrac{m_{p,0} c_{p,p,0} \left(T_{p}^{psy}-T_{p,0}\right)}{-\dot{Q}_{r,0} - L_{v,0} \dot{m}_{r,0}},
\end{align}
where $T_{p}^{psy}$ is the wet-bulb temperature and the $0$ subscript signifies the terms evaluated at the initial condition.  
Considering $m_{p,0} \sim d_{p,0}^3$, $\dot{Q}_{r,0} \sim d_{p,0}$, and $\dot{m}_{r,0} \sim d_{p,0}$, this heat-up time scale scales with the initial diameter as: $\tau_{p,heat} \sim d_{p,0}^2$ just like the evaporation time scale $\tau_{p,evap}$.

\begin{figure}[h!]
    \centering
	
	\resizebox{0.95\textwidth}{!}{
	\begin{tikzpicture}
	           
	   \node[above right] at (0,0) {
	       \includegraphics[width= 10 cm]{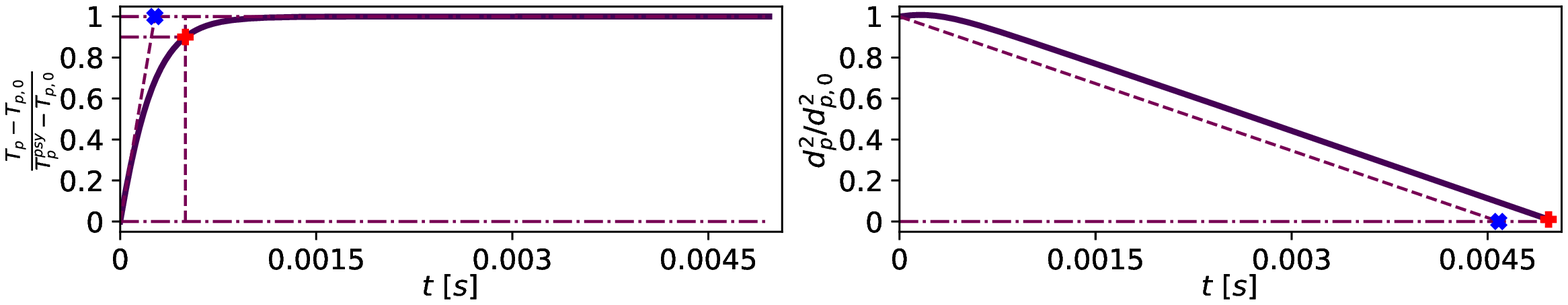}
	   };
	   
	   \node[above right] at (1.7,0.8) {
	       \includegraphics[width= 2.4 cm]{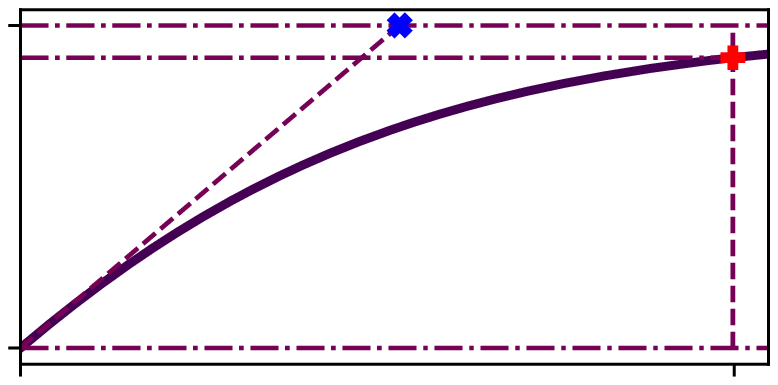}
	   };

	   \draw[very thick,-latex] (1.9,2.35) -- (3.06,2.35) node[midway,above,yshift=-0.1cm] {$\tau_{p,heat}$};
	   
	   \draw[very thick,-latex] (1.9,2.8) -- (4.07,2.8) node[midway,above,yshift=-0.1cm] {$\tau_{p,T90\%}$};

	   \draw (1.9,1.9) -- (1.9,2.9);
	   \draw (3.06,1.9) -- (3.06,2.45);
	   \draw (4.07,1.8) -- (4.07,2.9);

	   \draw[very thick,-latex] (5.88,2.35) -- (9.665,2.35) node[midway,above,yshift=-0.1cm] {$\tau_{p,evap}$};
	   
	   \draw[very thick,-latex] (5.88,2.8) -- (9.975,2.8) node[midway,above,yshift=-0.1cm] {$\tau_{p,tot}$};
	   
	   \draw (5.88,1.9) -- (5.88,2.9);
	   \draw (9.665,0.75) -- (9.665,2.45);
	   \draw (9.975,0.75) -- (9.975,2.9);

	\end{tikzpicture}
	}
	
    \caption{Illustration of the time scale estimations for the heat-up period and droplet lifetime. The blue x markers indicate the estimates, while red cross markers provide a reference based on the numerical simulation of the evaporation process. The heat up timescale is marked on a magnified plot for clarity.
    This reference case shows the evolution of an n-heptane droplet using Bird's correction with an initial diameter of $d_{p,0} = 50 \ \mathrm{\mu m}$, and initial temperature difference of $T_{p}^{psy}-T_{p,0} = 40 \ \mathrm{K}$. The droplet is stationary ($Re_p=0$) and the seen conditions are $T_s = 1500 \ \mathrm{K}$ and $Y_{f,s} = 0$.}
    \label{fig:time_scale_illustrations}
\end{figure}

Figure~\ref{fig:time_scale_illustrations} illustrates the two time scales: $\tau_{p,heat}$ and $\tau_{p,evap}$. 
The scales are plotted over the simulated evolution of an 
n-heptane droplet using Bird's correction with an initial diameter of $d_{p,0} = 50 \ \mathrm{\mu m}$, and initial temperature difference of $T_{p}^{psy}-T_{p,0} = 40 \ \mathrm{K}$ in air. 
The droplet is stationary ($Re_p=0$) and the seen conditions are $T_s = 1500 \ \mathrm{K}$ and $Y_{f,s} = 0$.
Figure~\ref{fig:time_scale_illustrations} also shows two time scales of the simulated droplet evolution. 
A heat-up time scale $\tau_{p,T90\%}$ is defined as the time when the droplet temperature has completed 90\% of the change between the initial temperature $T_{p,0}$, 
and the psychrometric wet-bulb temperature $T_{p}^{psy}$. 
And the lifetime of the droplet $\tau_{p,tot}$ is defined as the time it takes to reach $0.1\%$ of the initial droplet mass. Note, that the simulation is stopped at this point. 
 
Such simulations are executed over a wide range of parameters to study the model behavior and compare the estimates $\tau_{p,heat}$ and $\tau_{p,evap}$ to their simulated counterparts $\tau_{p,T90\%}$ and $\tau_{p,tot}$.
The four fuels: OME1, n-heptane, n-decane, and n-dodecane are studied using the proposed models: diffusion only (D/D), classical (S/D), Bird's correction (B), Abramzon-Sirignano (AS).
The varied parameters are the initial droplet diameter
$d_{p,0} \in \{ 0.5, 5, 50, 500 \} \ \mathrm{\mu m}$,
the difference between the psychrometric wet-bulb temperature and the initial temperature: $T_{p}^{psy}-T_{p,0} \in \{ 5,10,20,40 \} \ \mathrm{K}$
the Reynolds number $Re_p \in \{0,1,10,100,1000\}$,
and the seen temperature 
$T_s \in \{500, 750, 1000, 1250, 1500, 1750, 2000 \} \ \mathrm{K}$. 
For simplicity the seen vapor mass fraction is kept constant zero.

\begin{figure}[h!]
    \centering
	\includegraphics[width= 0.95 \textwidth]{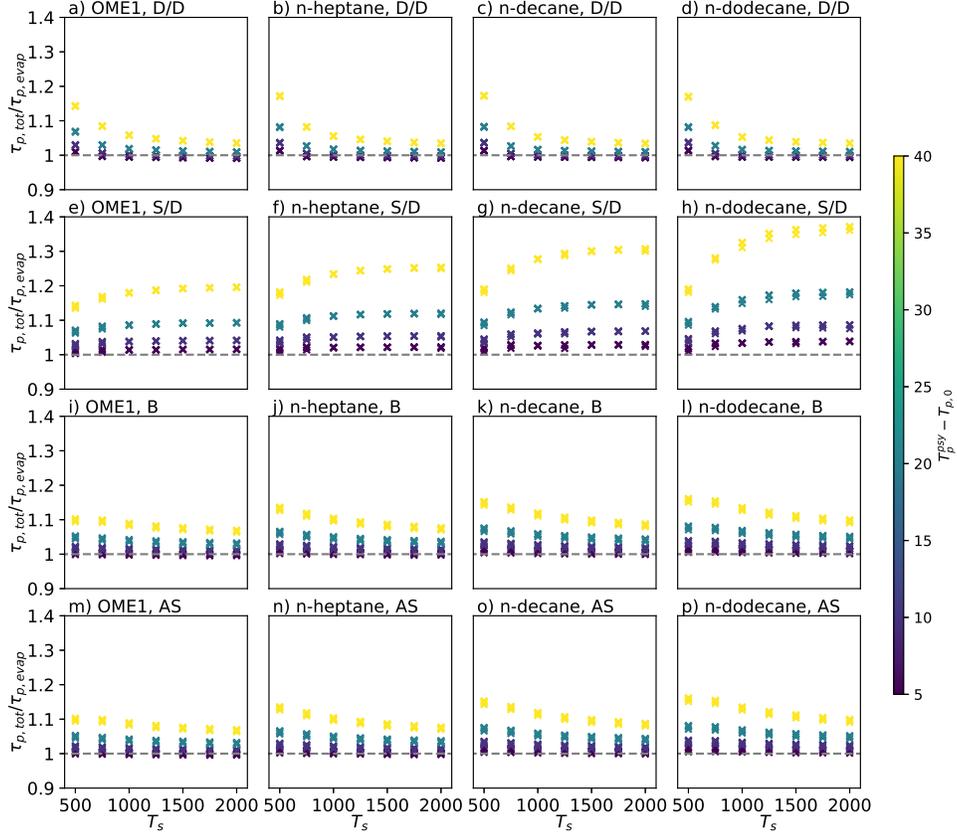}
    \caption{Comparison of the evaporation timescale estimate $\tau_{p,evap}$ and the time necessary to evaporate $99.9\%$ of the initial droplet mass in simulations. The ratio of the two time scales is assessed as function of the seen gas temperature under different initial temperatures marked by the color scheme, under different constant Reynolds numbers, and with different initial droplet sizes. \mynormaltext{The droplet size and the Reynolds number are not indicated as there is insignificant dependence on these parameters.}}
    \label{fig:evap_estimate}
\end{figure}

Figure~\ref{fig:evap_estimate} illustrates the total simulation time $\tau_{p,tot}$, compared to the estimate assuming the droplet evaporates under wet-bulb conditions $\tau_{p,evap}$. 
\mynormaltext{The ratio $\tau_{p,tot} / \tau_{p,evap}$ is displayed as a function of the seen gas temperature $T_s$. The color scheme indicates the initial temperature difference $T_{p}^{psy}-T_{p,0}$.
The effect of droplet Reynolds number $Re_p$ on this ratio is negligible and only $Re_p=0$ is displayed here.
Likewise, the different initial droplet diameters $d_{p,0}$ are not distinguished as the symbols are completely overlapping.}

As Fig.~\ref{fig:time_scale_illustrations} shows, this ratio is an indication of what fraction of the droplet lifetime is spent with heat-up\mynormaltext{, since $\tau_{p,tot}-\tau_{p,evap}$ is the additional time the droplet spends with reduced evaporation rate due to temperatures lower than the wet-bulb temperature}. 
The initial droplet diameter $d_{p,0}$ has no effect on this property, the droplet Reynolds number $Re_p$ has limited influence in the Abramzon-Sirignano model (AS) only\mynormaltext{, however, it is too small to visualize.}
There is a slight dependence on the seen gas temperature $T_s$, and most of the variation can be attributed to the difference between the initial temperature and the wet-bulb temperature. 
As expected, the $\tau_{p,tot} / \tau_{p,evap}$ ratio decreases as the initial droplet temperature approaches the wet-bulb temperature, since the heat-up period diminishes.

Bird's correction (B) and the Ambramzon-Sirignano model (AS) show similar trends 
\mynormaltext{even at non-zero Reynolds numbers not shown here,} as the presented ratio decreases with the increase of the seen gas temperature $T_s$, indicating that the relative importance of the heat-up period diminishes in high temperature environments. 
The classical model (S/D) shows an opposite trend, that is better understood observing the evaporation rate constants of Fig.~\ref{fig:validation}. One may observe, that the evaporation rate constant increases faster than linear as function of the seen gas temperature. 
Meanwhile the heat up time scale is relatively linear as a function of $T_s$, thus the classical model (S/D) predicts higher and higher fractions of time spent on the heat-up.

\mynormaltext{A similar comparison of the time scale estimate and the simulated time scale is studied for the heat-up period in \ref{sec:heat_up_time_scale}, comparing $\tau_{p,T90\%}$ to $\tau_{p,heat}$. 
In conclusion the heat-up is completed in $1.5$ to $2$ times the $\tau_{p,heat}$ estimate for Bird's correction and the Abramzon-Sirignano model, influenced mostly by the initial droplet temperature.
The analysis of \ref{sec:heat_up_time_scale} further emphasizes the inadequacy of the diffusion only and classical models. 
}

Overall, according to the analytical derivations of Section~\ref{sec:modelling}, the diffusion only model (D/D) neglects an important part of the physical phenomena involved in evaporation: Stefan flow, 
while the classical model (S/D) considers it wrongly.
The validation against the experimental data of Chauveau~et~al.~\cite{chauveau2019analysis} underlines this discrepancy as illustrated in Fig.~\ref{fig:validation}.
In the remaining part of the study, the diffusion only model (D/D) and the classical model (S/D) are disregarded, since their validity is limited to low temperature applications.  
Only the models correctly considering Stefan flow are analyzed below.

\subsubsection{Reynolds number effects in the Abramzon-Sirignano model (AS)}

As shown in Fig.~\ref{fig:AS_corr}, the Abramzon-Sirignano model (AS) introduces a  modification to Bird's correction (B), that only acts in case of finite film thickness, i.e.: in case of non-zero Reynolds number. The correction is limited to a maximum of $22\%$ reduction of the transfer rates in very high Reynolds numbers. 
\begin{figure}[h!]
    \centering
	\includegraphics[width= 0.95 \textwidth]{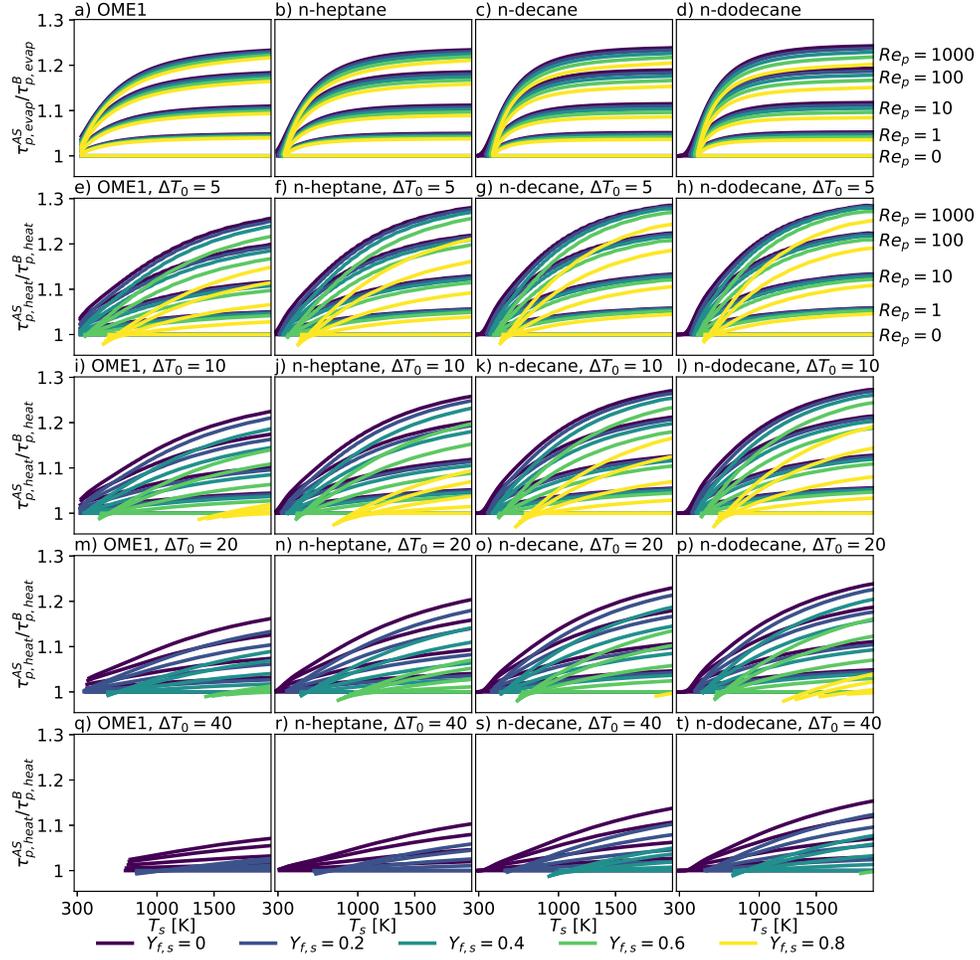}
    \caption{Comparison of the evaporation (a-d) and heat-up (e-t) timescales between the Abramzon-Sirignano model (AS) and Bird's correction (B). The evaporation time scales are compared under the wet-bulb conditions of the respective models, while the heat-up time scales are assessed using the same initial temperature: $T_{p,0} = T_{p}^{psy,B} - \Delta T_0$. }
    \label{fig:AS_B_compare_K}
\end{figure}

The ratio of time scale estimates of evaporation are shown in Fig.~\ref{fig:AS_B_compare_K}~a-d  and of heat-up at different initial temperatures in Fig.~\ref{fig:AS_B_compare_K}~e-t for the Abramzon-Sirignano model (AS) and Bird's correction (B) respectively.
As expected, the additional correction introduced by Abramzon and Sirignano~\cite{abramzon1989droplet} increases the time scales for the combination of high Reynolds numbers with high Spalding mass transfer numbers. 
The degree of time scale augmentation is within $30\%$ under the studied conditions. 
These values are approached only at high seen temperatures and high Reynolds numbers. 
For low temperature applications, the Abramzon-Sirignano model (AS) only affects the highly volatile OME1, the rest of the studied fuels is practically unaffected at seen gas temperatures of $300 \ \mathrm{K}$. 

The evaporation time scales displayed in Fig.~\ref{fig:AS_B_compare_K}~a-d are evaluated under the wet-bulb conditions according to \mynormaltext{$\tau_{p,evap} = \tfrac{ d_{p,0}^2}{ K }$}.
This ratio shows a clear growth as function of the Reynolds number $Re_p$, and the seen gas temperature $T_s$. 
The correction factor of Abramzon and Sirignano~\cite{abramzon1989droplet} saturates after a certain Spalding mass transfer numbers is reached, thus the $\tau_{p,evap}^{AS}/\tau_{p,evap}^{B}$ ratio similarly reaches a plateau with increasing seen gas temperatures.
The dependence on seen vapor mass fraction is minor.

The heat-up time scales are shown for different initial temperatures in Fig.~\ref{fig:AS_B_compare_K}~e-t where $\Delta T_0 = T_{p}^{psy,B}-T_{p,0}$ is the difference between the psychrometric wet-bulb temperature given by Bird's correction and the initial droplet temperature. 
Note, that this means that $T_{p,0}$ is the same for both Bird's correction (B) and the Abramzon-Sirignano model (AS) even though $T_{p}^{psy,B}$ and $T_{p}^{psy,AS}$ are slightly different. 
The effect on the heat-up timescale diminishes as $\Delta T_0$ increases and the droplets get further from the wet-bulb conditions, since the transfer rates are low at high $\Delta T_0$, thus only part of the heat-up process is really affected.
Comparing Fig.~\ref{fig:AS_B_compare_K}~a-d and Fig.~\ref{fig:AS_B_compare_K}~e-h, one can observe a difference between the behavior of mass and heat transfers. 
Overall, the correction of the heat transfer time scale is higher than that of the mass transfer time scale, 
as $\phi_m$ tends to be over unity at higher seen temperatures. 
The Spalding mass transfer number is limited to $B_M < 6$ under the studied conditions (see Fig.~\ref{fig:psych_wet_bulb_bird}), 
so $B_T^* > B_M$ and the correction of heat transfer can reach the maximum $28\%$ while that of mass transfer cannot (see Fig.~\ref{fig:AS_corr}).

In general, the Abramzon-Sirignano model (AS) introduces significant changes compared to Bird's correction (B) at high Reynolds numbers, and the effect is notable even at $Re_p = 1 .. 10$. 
Such droplet Reynolds numbers are typically sustained in a turbulent flow field, where the variability of the gas phase velocity and the inertia of the droplets keeps up a non-zero slip velocity.
Thus, the usage of the Abramzon-Sirignano model (AS) is recommended for spray combustion simulations.

\subsubsection{Langmuir-Knudsen model (LK1,LK2)}
\label{subsubsec:LK_single_drop}

The Langmuir-Knudsen models described in Section~\ref{sec:modelling} are rather particular in the sense, that these models introduce diameter dependence on quantities, that are independent of the diameter in all the other studied models. 
For this reason, the wet-bulb conditions are undefined as the droplets do not approach a specific equilibrium temperature during their lifetime as it is the case with the other studied models. 
Furthermore, as already illustrated in \mynormaltext{\ref{sec:limit_LK}}, the non-equilibrium mass transfer number $B_M^{neq}$ is limited depending on the droplet diameter. 
Thus applying the Langmuir-Knudsen models, the droplets may reach a minimum diameter $d_{p,min}$ in their lifetime, where the liquid droplet temperature approaches the boiling point (thus $B_M^{eq} \rightarrow \infty$) and energy conservation cannot be satisfied because the mass transfer is limited by $\max(B_M^{neq})$, similarly to the case of the diffusion only model (D/D).

A number of single droplet simulations were executed using the two different Langmuir-Knudsen models: LK1 and LK2. 
The chosen parameter set is similar as before: 
the initial droplet diameter is:
$d_{p,0} \in \{ 0.5, 2, 5, 20, 50, 500 \} \ \mathrm{\mu m}$,
the difference between the psychrometric wet-bulb temperature calculated with Bird's correction and the initial temperature is: $\Delta T_0 = T_{p}^{psy,B}-T_{p,0} \in \{ 0,40 \} \ \mathrm{K}$
the Reynolds number is: $Re_p \in \{0,10,1000\}$,
and the seen temperature is varied between $300 \ \mathrm{K}$ and $2000 \ \mathrm{K}$ with $100 \ \mathrm{K}$ steps. 
For simplicity the seen vapor mass fraction is again kept constant zero.
The simulations are run either until $99.9\%$ of the initial droplet mass is evaporated, or until the $B_M^{eq} > 10^5$ condition is satisfied indicating that $d_{p,min}$ is reached. 
\mynormaltext{Latter limit is further studied in \ref{sec:limit_LK}, showing that the minimum attainable diameter can be estimated as: $d_{p,min}^{LK1} \approx 2 L_K^B \phi_m^B \ln\left(1+10^5\right)$ for the non-iterative LK1 model. 
The limiting diameter of the iterative LK2 model follows the same trend, but the diameter is approximately half: $d_{p,min}^{LK2} \approx d_{p,min}^{LK1} / 2$. }

\begin{figure}[h!]
    \centering
	\includegraphics[width= 0.95 \textwidth]{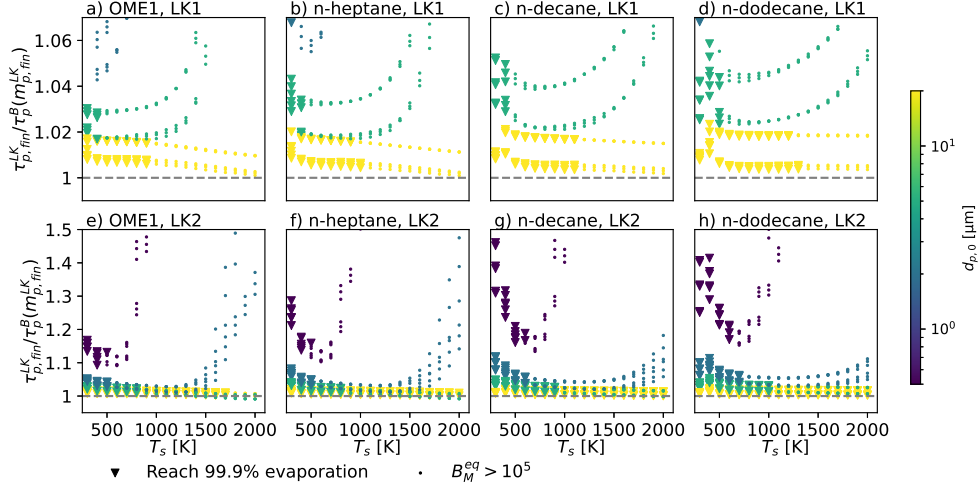}
    \caption{ Ratio of time taken till the final droplet mass is reached in case of Langmuir-Knudsen models (LK1 and LK2) and Bird's correction (B) as function of the seen gas temperature. The models are assessed under various seen gas temperatures, initial temperatures, Reynolds numbers, and initial droplet diameters. The initial diameter is indicated by the color scheme, while symbols indicate the cause of termination of the Langmuir-Knudsen simulations. }
    \label{fig:LK_time_ratio}
\end{figure}

The simulated cases are analyzed in  Fig.~\ref{fig:LK_time_ratio} presenting the time necessary to reach the final possible droplet mass 
normalized by the time necessary to reach the same mass using Bird's correction. 
I.e.: in case it is possible to evaporate $99.9\%$ of the initial droplet mass, the presented ratio is $\tau_{p,tot}^{LK} / \tau_{p,tot}^{B}$ (represented by triangles)\mynormaltext{, where $\tau_{p,tot}$ is defined in Fig.~\ref{fig:time_scale_illustrations}}. Otherwise, if the droplet evaporation cannot be completed, then the denominator of the ratio is interpolated from the complete simulation using Bird's correction corresponding to the same mass (represented by dots).
This figure quantifies the importance of using the Langmuir-Knudsen models, as the non-equilibrium models only deviate from Bird's correction for very small droplet sizes. 
Above an initial droplet size of $20 \ \mathrm{\mu m}$ ,this deviation is completely insignificant, and is omitted here. I.e.: $99.9\%$ of the droplet mass can evaporate, without any notable non-equilibrium effects. 

As the initial droplet size decreases, the non-equilibrium models (LK1,LK2) become more important, however the diameter limitation \mynormaltext{discussed in Appendix C} restricts their applicability to \mynormaltext{low} seen temperatures as the evaporation is terminated prematurely. 
In general, the non-iterative LK1 model suffers from this limitation \mynormaltext{to a greater extent}. 
The $d_{p,0} = 20 \ \mathrm{\mu m}$ droplets can complete their evaporation under most studied conditions, but smaller droplets cannot. 
Overall, the effects of the LK1 model are either small, because of the larger droplet size, or the effects get severe enough to impede full evaporation.

The iterative solution of the non-equilibrium conditions (LK2) is much less restricted in terms of $d_{p,min}$, thus even droplets of $d_{p,0} = 0.5 \ \mathrm{\mu m}$ can be successfully simulated under certain seen temperatures. 
For this reason the observed effect can be much larger in the cases that complete the evaporation (Fig.~\ref{fig:LK_time_ratio} e-h triangles).
In general, the effect increases with larger and less volatile hydrocarbons. 
In most fuels, the effect monotonously decreases with increasing seen gas temperature, except in the case of n-dodecane. 
This characteristic change of behavior is related to the extremely low volatility of n-dodecane at $T_s = 300 \ \mathrm{K}$ that can be observed in the other analysis of the present work.

\begin{figure}[h!]
    \centering
	\includegraphics[width= 0.95 \textwidth]{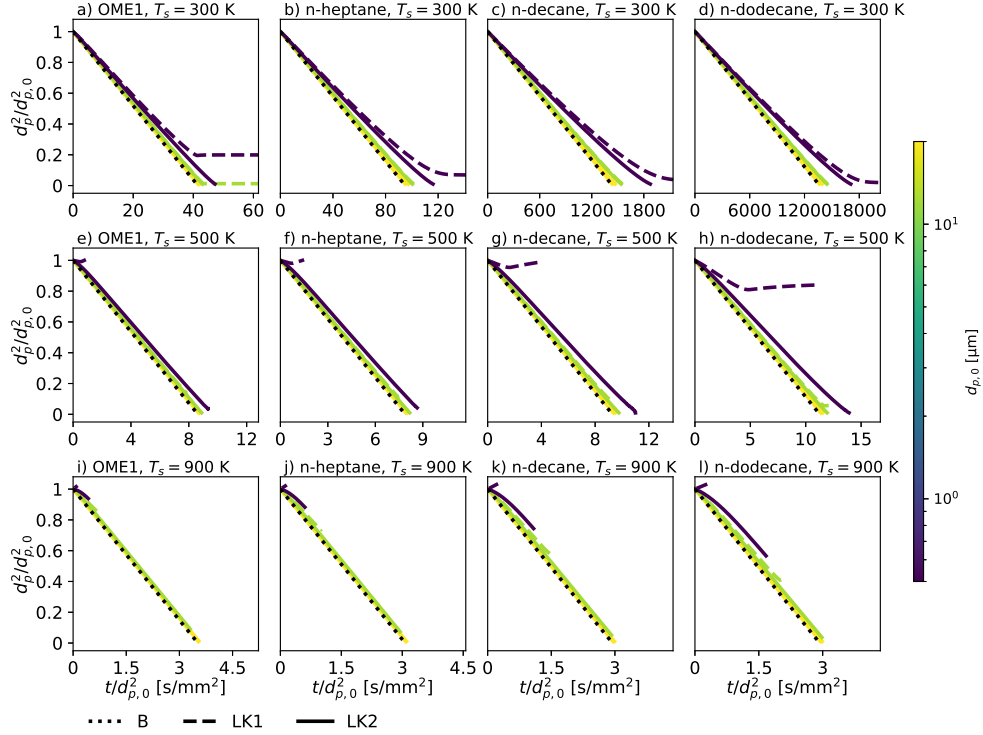}
    \caption{ Evolution of droplet surface in time using the non-iterative (LK1) and iterative (LK2) Langmuir-Knudsen models under various seen gas temperatures. The Reynolds number and seen vapor mass fraction are zero. Three initial diameters are indicated by the color scheme in $d_{p,0} \in \{ 0.5, 2, 5\} \ \mathrm{\mu m}$, while the initial droplet temperature is given by the wet-bulb conditions using Bird's correction. The equilibrium solution using Bird's correction (B) is indicated for reference. }
    \label{fig:LK_examples}
\end{figure}

The behavior of the Langmuir-Knudsen models is further analyzed in Fig.~\ref{fig:LK_examples}. 
The figure shows $d_{p}^2/d_{p,0}^2$ as function of $t/d_{p,0}^2$, for stationary droplets ($Re_p=0$) under different seen gas temperature, and zero seen vapor mass fraction. 
The scaling of the coordinates, makes the plot independent of initial droplet size in case of Bird's correction, as the evaporation constant is only a function of the seen conditions and material properties. 
Thus, Fig.~\ref{fig:LK_examples} can highlight the differences introduced by the non-equilibrium models. 
The figure provides examples of simulations using the iterative and non-iterative models for three different initial droplet diameters. 
In this scale, the effect appears quite insignificant even for these small droplets except for the sub-micron case of $d_{p,0} = 0.5 \ \mathrm{\mu m}$. 

The evolution of the droplet size highlights a key issue of the non iterative method (LK1) related to the phenomena shown in discussed in \mynormaltext{\ref{sec:limit_LK}}: the non-equilibrium mass transfer number $B_M^{neq}$ has a local maximum in droplet temperature $T_p$ then drops to zero and this zero-crossing is rather far from the boiling point in case of sub-micron droplet diameters (Fig.~\ref{fig:BM_Tp_LK} i-l). 
Thus the LK1 model can completely impede the evaporation process before the highest temperatures are reached, resulting in cases like the one presented in Fig.~\ref{fig:LK_examples} h) for $d_{p,0} = 0.5 \ \mathrm{\mu m}$ where the droplet temperature keeps increasing even though the evaporation is over. 
The LK2 model does not show this behavior, as it asymptotically approaches a maximum $B_M^{neq}$ as the droplet temperature increases (Fig.~\ref{fig:BM_Tp_LK}).
However, this model is also limited by a minimum possible diameter. 

Overall, the applicability of the Langmuir-Knudsen models is limited on two fronts.
On one side, if the initial droplet diameter is too large, the models have barely any effect on the major part of the evaporation process ($99.9\%$ of the initial mass can be evaporated without any significant effect.)
On the other side, if the initial droplet diameter is too small, the models are limited by the minimum achievable diameter. 
Latter limitation increases with the seen gas temperature, thus the Langmuir-Knudsen models can only be used in low temperature studies.
Furthermore, experimental evidence is lacking for assessing the performance of these models, as sub-micron measurements are not yet possible. 
State of the art measurements can study droplets of $d_{p,0} = \mathcal{O}\left(100\right) \ \mathrm{\mu m}$. \cite{chauveau2019analysis}

\section{Concluding Remarks}\label{Conclusions}

The fundamentals of analytical heat and mass transfer sub-models have been reviewed in the context of film theory for spherical droplets. 
The combination of these two sub-models yield widely used evaporation models under the infinite conductivity assumption, that describe the evolution of droplet size and temperature at given far-field ("seen") gas phase conditions. 
The family of models considering 
Stefan flow in both heat and mass transfer stand out in terms of performance among the studied options, 
namely the model denoted as Bird's correction (B), the Abramzon-Sirignano model (AS) and the Langmuir-Knudsen model (LK).
The two other studied models either ignore Stefan flow as in the case of diffusion only model (D/D), or partially ignore it in case of the classical model (S/D). 
Both resulting in nonphysical behavior in high temperature environments.
Unfortunately many of these models are validated at low temperature conditions where all of them behave very similarly. 

It must be noted, that Bird~et~al.~\cite[§19.4,§22.8]{bird1960transport} originally derived the heat transfer correction term for evaporation or condensation over a flat plate. 
This correction is expressed in terms of non-dimensional numbers ($\beta$) and 
it is used on spherical cases such as evaporating droplets. 
In the present work, the correction is derived from first principles to spherical coordinates, 
yielding the surprising conclusion, that Bird's correction is indeed the same for droplets and flat plates despite the fundamental differences in configuration.

The evaporation characteristics of four different pure compounds: OME1, n-heptane, n-decane, and n-dodecane are studied using the \mynormaltext{aforementioned models}.
These fuels differ in terms of volatility, that causes the most variation between their behavior. 
The difference is most striking in low temperature environment, where n-dodecane behaves radically different from the more volatile fuels.
\mynormaltext{The evaporation of single droplets of these fuels is numerically investigated under an extensive range of conditions from ambient to flame-like environments.}
It is found, that the initial heat-up process can extend the droplet lifetime by $\sim10\%$ if the initial temperature is sufficiently far from the wet-bulb conditions.

Finally, the additional considerations of non-equilibrium thermodynamics~\cite{miller1998evaluation} and interaction between the mean flow and the Stefan flow~\cite{abramzon1989droplet} 
are evaluated. 
It is found that the Langmuir-Knudsen model needs an iterative process, to correctly evaluate the non-equilibrium vapor pressures. 
Even with this iterative solution, the application of this model shall be limited to low temperature evaporation of sub-micron droplets, where the computation is not limited neither by the inherent instability of the model, nor by its negligible effect compared to Bird's correction. 
Thus, the Langmuir-Knudsen models are not suitable for combustion simulations.
The Reynolds number effects considered by the Abramzon-Sirignano model are found to be significant even at relatively low Reynolds numbers. 
The authors consider this model to be the state of art in Lagrangian fuel spray modeling under the conditions of the present work\mynormaltext{, and it can be used with confidence for liquid fuel combustion applications.}

\section*{Conflict of interest}

The authors have no conflict of interest to declare.

\section*{Acknowledgments}

We acknowledge the funding received 
through the ESTiMatE project from the Clean Sky 2 Joint Undertaking under the European Union’s Horizon 2020 research and innovation programme under grant agreement No 821418
and
through the Spanish Ministry of Economy and Competitiveness 
in the frame of the AHEAD project (PID2020-118387RB-C33) 
.

\appendix

\section{Wet-bulb conditions of the diffusion only model (D/D)}
\label{sec:wet_bulb_DD}

\mynormaltext{For} the diffusion only model (D/D), the psychrometric wet-bulb conditions are given by: 
\begin{align}
    \label{eq:PsychroWetBulb_DD}
    c_{p,m} \left(T_s-T_p^{psy,D/D}\right) &= \dfrac{1}{Le_m} \dfrac{Sh_{m,0}}{Nu_{m,0}} \left(Y_{f,i}^{psy,D/D}-Y_{f,s}\right) L_v.
\end{align}
The far-field temperature $T_s$, and vapor mass fraction $Y_{f,s}$ are boundary conditions of the problem, while the wet-bulb temperature $T_p^{psy,D/D}$ is the unknown, \mynormaltext{and} the interface vapor mass fraction $Y_{f,i}^{psy,D/D}$ is a monotonous increasing function \mynormaltext{in temperature} up to the boiling point of the liquid.  
At atmospheric pressure, far from the critical point, the latent heat of vaporization $L_v$ is only weakly dependent on the droplet temperature. 
The other coefficients: $c_{p,vap,m}$ and $\phi_m$ are also constrained to a range of finite values.
Thus, Eq.~\eqref{eq:PsychroWetBulb_DD} only has a solution for a constrained range of far-field temperatures, unlike in the case of thermodynamic wet-bulb conditions. 
This highlights the limitation of neglecting Stefan flow in the evaporation model. 

\begin{figure}[h!]
    \centering
	\includegraphics[width= 0.95 \textwidth]{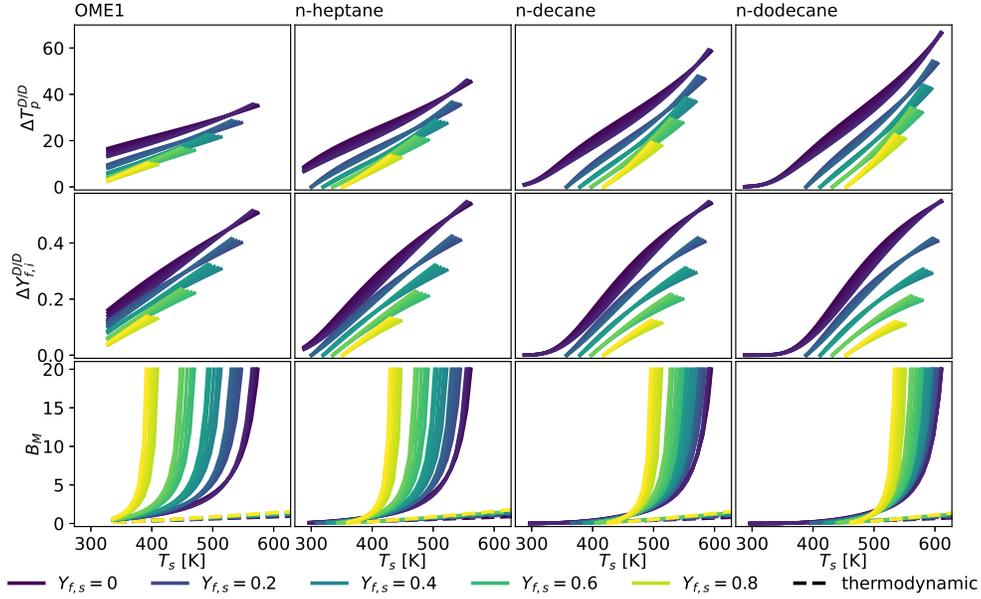}
    \caption{Comparison of psychrometric and thermodynamic wet-bulb conditions of OME1, n-heptane, n-decane, and n-dodecane at atmospheric pressure with air as bath gas  according to the diffusion only model (D/D): Eq.~\eqref{eq:PsychroWetBulb_DD}. The difference in wet-bulb temperatures and the corresponding vapor mass fractions are $\Delta T_p^{D/D} = T_p^{psy,D/D} - T_p^{th}$, and $\Delta Y_{f,i}^{D/D} = Y_{f,i}^{psy,D/D} - Y_{f,i}^{th}$.}
    \label{fig:psych_wet_bulb_DD}
\end{figure}

Figure~\ref{fig:psych_wet_bulb_DD} shows the difference between the thermodynamic wet-bulb conditions and the wet-bulb conditions given by Eq.~\eqref{eq:PsychroWetBulb_DD} for the diffusion only model where the differences in wet-bulb temperatures and the corresponding vapor mass fractions are $\Delta T_p^{D/D} = T_p^{psy,D/D} - T_p^{th}$, and $\Delta Y_{f,i}^{D/D} = Y_{f,i}^{psy,D/D} - Y_{f,i}^{th}$ respectively. 
The curves corresponding to equilibrium states are presented as a function of the seen gas temperature $T_s$, and parametrized by the seen gas vapor mass fraction $Y_{f,s}$, and the Reynolds number $Re_m$. 
This latter dependence corresponds to Reynolds numbers of $Re_m \in \{0,1,10,100,1000\}$, the legend omits this dependence for simplicity, as the equilibrium conditions are rather insensitive to the Reynolds number in this exhaustive range.
Nonetheless to interpret the variation: lighter colors correspond to higher Reynolds numbers.

The vicinity of the saturation condition is illustrated best by the Spalding mass transfer number $B_M$.
It is evident, 
that in case of the diffusion only model (D/D), 
the wet-bulb conditions are only found below a certain seen gas temperature. 
In Fig.~\ref{fig:psych_wet_bulb_DD}, the \mynormaltext{wet-bulb calculations are arbitrarily cut-off } where $B_M=20$\mynormaltext{, thus the maximum displayed seen temperature is:}
$\max \left( T_s \right)^{D/D} = T_{p |B_M=20} + \tfrac{\phi_m}{c_{p,vap,m}} L_v \left( Y_{f,i |B_M=20}-Y_{f,s} \right)$. 
After this limit, the Spalding transfer number keeps approaching infinity, while the change in $\max \left( T_s \right)^{D/D}$ is \mynormaltext{small}. 
The limiting values \mynormaltext{on Fig.~\ref{fig:psych_wet_bulb_DD}} for $Y_{f,s} = 0$ and $Re_m = 0$ are:
$574.0 \ \mathrm{K}$, $561.9 \ \mathrm{K}$, $594.7 \ \mathrm{K}$, $610.9 \ \mathrm{K}$, for OME1, n-heptane, n-decane, and n-dodecane respectively. 
\mynormaltext{If a cut-off point of $B_M=200$ was chosen, the limiting seen temperatures would be approximately $20 \ \mathrm{K}$ higher.}
However it does not mean, that the effect of Stefan flow could be neglected below this limit. 
The application of this model should be limited to low temperature, however choosing it over the other models presented here is only justified by its computational simplicity. 
Considering the operations needed to evaluate the mean gas properties, the material properties of the liquid, and the phase change properties, this advantage is negligible compared to Bird's correction (B) (that do not need iterative methods to determine the rate of evaporation).
Thus the authors recommend avoiding the usage of the diffusion only model (D/D) \mynormaltext{altogether, especially in combustion applications}.

\section{Wet-bulb conditions of the classical model (S/D)}
\label{sec:wet_bulb_SD}

\begin{figure}[h!]
    \centering
	\includegraphics[width= 0.95 \textwidth]{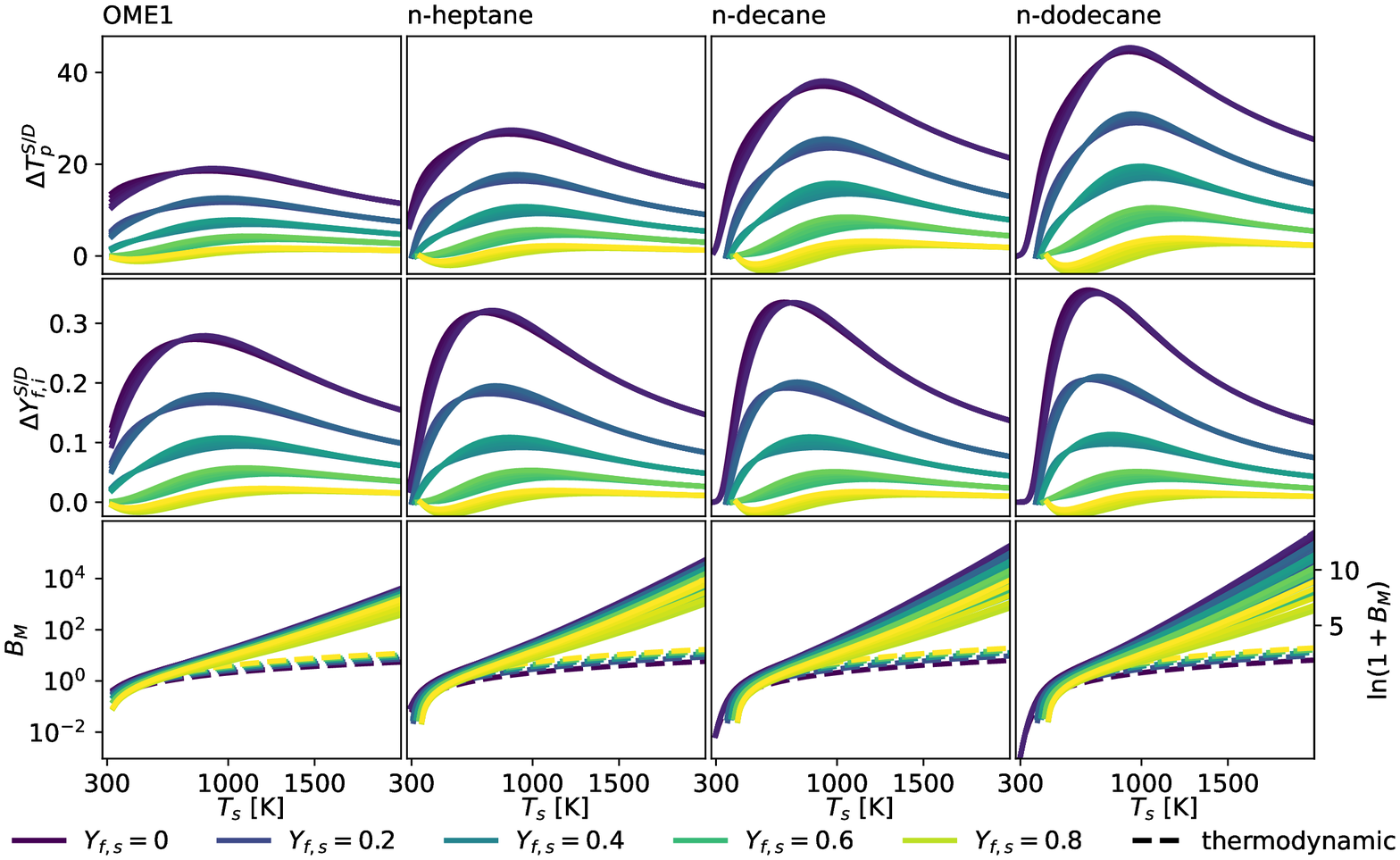}
    \caption{Comparison of psychrometric and thermodynamic wet-bulb conditions of OME1, n-heptane, n-decane, and n-dodecane at atmospheric pressure with air as bath gas according to the classical evaporation model (S/D):  Eq.~\eqref{eq:PsychroWetBulbBT_SD}. 
    The difference in wet-bulb temperatures and the corresponding vapor mass fractions are $\Delta T_p^{S/D} = T_p^{psy,S/D} - T_p^{th}$, and $\Delta Y_{f,i}^{S/D} = Y_{f,i}^{psy,S/D} - Y_{f,i}^{th}$.}
    \label{fig:psych_wet_bulb}
\end{figure}

In case of the classical evaporation model (S/D), the wet-bulb conditions are given by: 
\begin{align}
    \label{eq:PsychroWetBulbBT_SD}
    c_{p,vap,m} \left(T_s-T_p^{psy,S/D}\right) &= \ln\left(1+B_T^{psy,S/D}\right) L_v.
\end{align}
The differences between the psychrometric and thermodynamic wet-bulb conditions are illustrated in Fig.~\ref{fig:psych_wet_bulb} through the difference in wet-bulb temperature and vapor mass fraction\mynormaltext{: $\Delta T_p^{S/D} = T_p^{psy,S/D} - T_p^{th}$, and $\Delta Y_{f,i}^{S/D} = Y_{f,i}^{psy,S/D} - Y_{f,i}^{th}$ respectively}. 
 
The classical model yields an equilibrium state at all studied conditions, since $\ln\left(1+B_T^{psy,S/D}\right)$ is not limited as the droplet temperature approaches the boiling point. 
The equilibrium conditions are rather insensitive to the Reynolds number. 
The highest difference between the thermodynamic and psychrometric conditions is attained with dry air ($Y_{f,s}=0$). 

The presence of the logarithmic term causes the main difference between this model and the others $\left(B_M \ne \ln\left(1+B_M\right) \right)$.
To illustrate this, $\ln\left(1+B_M\right)$ is shown on the right axis of Fig.~\ref{fig:psych_wet_bulb}. 
To maintain equilibrium temperatures, $\ln\left(1+B_M^{psy}\right)$ has to be in the same order of magnitude as $B_M^{th}$ in Fig.~\ref{fig:therm_wet_bulb} for high temperature seen gas.
This results in significantly higher droplet temperatures, and surface vapor mass fractions, although not as high as with the diffusion only model. 

The main issue with the classical model (S/D) is the disparity \mynormaltext{caused by} considering Stefan flow in the mass transfer, but not in the heat transfer.

\section{Limit of applicability of the Langmuir-Knudsen models (LK)}
\label{sec:limit_LK}

\begin{figure}[h!]
    \centering
	\includegraphics[width= 0.97 \textwidth]{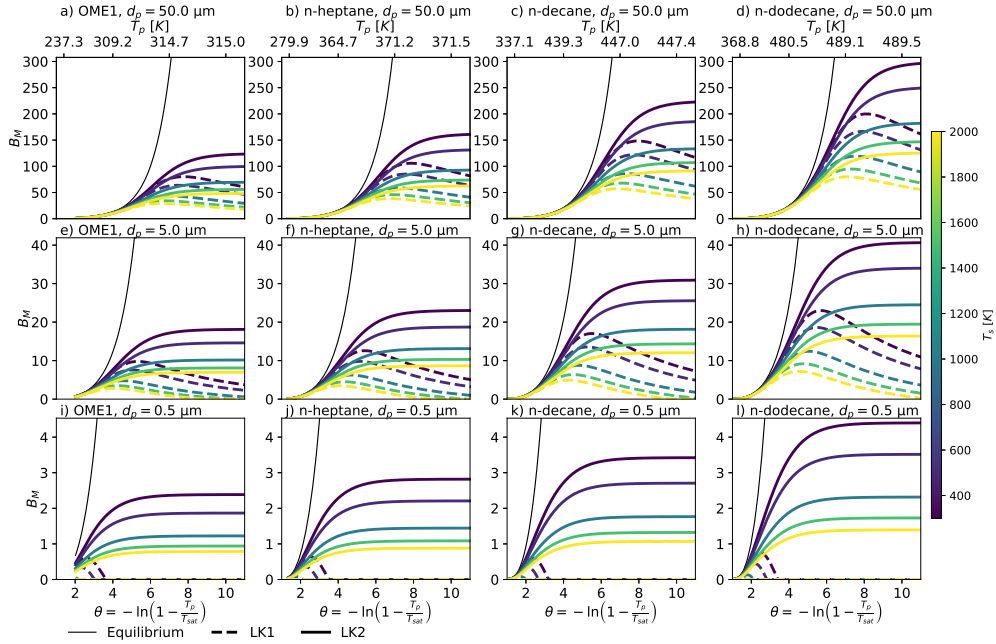}
    \caption{Non-equilibrium Spalding mass transfer numbers given by model LK1 and LK2 as function of droplet temperature for various liquids, at three different droplet diameters: a,b,c,d) $50 \ \mu\mathrm{ m}$, e,f,g,h) $5 \ \mu\mathrm{ m}$, and i,j,k,l) $0.5 \ \mu\mathrm{ m}$, and at 5 different seen temperatures: $300 \ \mathrm{K}$, $500 \ \mathrm{K}$, $1000 \ \mathrm{K}$, $1500 \ \mathrm{K}$, and  $2000 \ \mathrm{K}$. The seen gas is dry air at atmospheric pressure.}
    \label{fig:BM_Tp_LK}
\end{figure}

In case of the Langmuir-Knudsen models (LK), finding the wet-bulb temperature becomes more complex, as size-dependence interferes with the results.
To illustrate this, the non-equilibrium Spalding mass transfer numbers of model LK1 and LK2 are presented in Fig.~\ref{fig:BM_Tp_LK} as function of the droplet temperature at different seen temperatures, and different droplet diameters. Note that a coordinate transformation is applied in the droplet temperature: $\theta = - \ln\left(1-\tfrac{T_p}{T_{sat}}\right)$ to highlight the behavior near the boiling point.
Also, this figure differs from the previously shown examples in the sense, that the droplet is not at its steady state temperature. 
Generally, the mass transfer resembles the equilibrium solution better at higher diameters and low droplet temperatures.

The model without the iterative solution of $X_{f,i}^{neq}$ (LK1) is illustrated with dashed lines. 
As the droplet temperature increases the LK1 model shows local maxima in the mass transfer number, and even drops below 0 with the further increase of the droplet temperature. Not shown in the graph, this results in nonphysical cases where the non-equilibrium interface vapor mass fraction is negative 
\mynormaltext{in the further testing of the model, these conditions are omitted, and $X_{f,i}^{neq}\ge0$ is imposed}.
For the $50 \ \mu\mathrm{m}$ droplets (Fig.~\ref{fig:BM_Tp_LK} a,b,c,d) this shortcoming only takes effect within $1 \ \mathrm{K}$ of the boiling point, however as the droplets evaporate, larger and larger portions of the range of viable droplet temperatures is affected. 
In conclusion, the LK1 model \mynormaltext{cannot be recommended for droplets evaporating at high temperatures like combustion applications}, as it completely eliminates mass transfer at high droplet temperatures.

As the solid curves of Fig.~\ref{fig:BM_Tp_LK} illustrate, the LK2 model also limits the mass transfer numbers to a maximum, but $B_M^{neq}$ stays injective respect to the droplet temperature. 
The maximum attainable mass transfer number is a function of droplet size, seen temperature and gas composition. $B_M^{neq}$ decreases with the seen temperature as the "1/3 law" gives higher Knudsen layer thicknesses.  
Overall, the non-equilibrium mass transfer numbers still approach 0 as the droplet size decreases, but the non-physical local maxima and negative mass fractions are avoided with the iterative solution of $\beta^{neq} = \phi_m \ln\left(1+B_M^{neq}\right)$. 
$B_M^{neq}$ also decreases with increasing seen vapor mass fraction, which is not shown here for simplicity.

In conclusion, the Langmuir-Knudsen model may not provide a solution for Eq.~\eqref{eq:PsychroWetBulbLK}, 
as $B_T^{neq,psy,LK}$ is bounded since $B_M^{neq}$ is bounded as shown in Fig.~\ref{fig:BM_Tp_LK}.
On the left hand side of Eq.~\eqref{eq:PsychroWetBulbLK} $T_p^{psy,LK}$ is bounded by the boiling point, but $T_s$ is unbounded, thus the possible equilibrium states are limited just like in the case of the diffusion only model (D/D).
In case a sufficiently high $B_T^{neq,psy,LK}$ cannot be provided, there is no equilibrium state, however it does not mean the model is invalid, as the temperature takes a finite time to relax towards new equilibrium states. 
As Fig.~\ref{fig:BM_Tp_LK} illustrates, the range of feasible Spalding mass transfer numbers can accommodate the necessary values for droplets of $d_p=50 \ \mathrm{\mu m}$, 
since the mass transfer number only plateaus between 50 and 150 even for a seen temperature of $2000 \ \mathrm{K}$, 
but under these seen conditions, $B_M < 6$ is sufficient to keep a equilibrium temperature (Fig.~\ref{fig:psych_wet_bulb_bird}). 
The range of mass transfer number necessary for equilibrium is only unattainable for very small droplets of $d_p = \mathcal{O}\left(0.1 \ \mathrm{\mu m}\right) .. \mathcal{O}\left(1 \ \mathrm{\mu m}\right)$.
As demonstrated below, in practice the Langmuir-Knudsen model (LK2) largely behaves similarly to Bird's correction (B) for droplets that start the evaporation in the $d_p=\mathcal{O}\left(10 \ \mathrm{\mu m}\right)$ range, with a small interval near the end of the droplets lifetime, where the non-equilibrium effects slow down the mass transfer and the droplet temperature can quickly rise.

\begin{figure}[h!]
    \centering
	\includegraphics[width= 0.95 \textwidth]{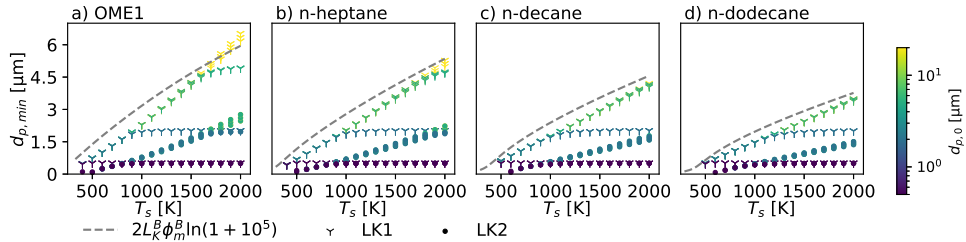}
    \caption{ Minimum diameter of applicability of the Langmuir-Knudsen models (LK1 and LK2) as function of the seen gas temperature identified as the diameter where $B_M > 10^5$ is reached in single droplet simulations. The models are assessed under various seen gas temperatures, initial temperatures, Reynolds numbers, and initial droplet diameters. }
    \label{fig:LK_min_diam}
\end{figure}

\mynormaltext{
The single droplet simulation data of subsection~\ref{subsubsec:LK_single_drop} is used below to further demonstrate the limitations of the Langmuir-Knudsen models.
In these simulations the droplets may reach a minimum diameter $d_{p,min}$ in their lifetime, where the liquid droplet temperature approaches the boiling point. 
}
Figure~\ref{fig:LK_min_diam} presents $d_{p,min}$ as the final diameter in the simulation cases where $B_M^{eq} > 10^5$ is reached. 
Such conditions are only observed starting from the initial diameter of 
$d_{p,0} \in \{ 0.5, 2, 5, 20\} \ \mathrm{\mu m}$. 
In the studied parameter range $d_{p,min}$ appears to be independent of the initial droplet temperature $T_{p,0}$, and there is only a slight dependence on the  Reynolds number $Re_p$. 
Furthermore, the initial droplet diameter $d_{p,0}$ only affects the minimum possible diameter, if the energy balance is unsatisfied almost instantly after the start of the simulation and $d_{p,min} \approx d_{p,0}$. 
Thus the determining factor of $d_{p,min}$ for a given fuel is the seen gas temperature $T_s$.
The minimum diameter of the LK1 model can be estimated well, by solving for $d_p$ assuming $B_M^{eq} = 10^5$ and $X_{f,i}^{neq} = 0$ and using material properties from the wet-bulb conditions of Bird's correction (B) as indicated by the dashed lines in Fig.~\ref{fig:LK_min_diam}: $d_{p,min}^{LK1} \approx 2 L_K^B \phi_m^B \ln\left(1+10^5\right)$.
As expected, the iterative process of the LK2 model extends its applicability range, and produces a minimum diameter less than half of the LK1 model.

\section{Heat-up time scale}
\label{sec:heat_up_time_scale}

\begin{figure}[h!]
    \centering
	\includegraphics[width= 0.95 \textwidth]{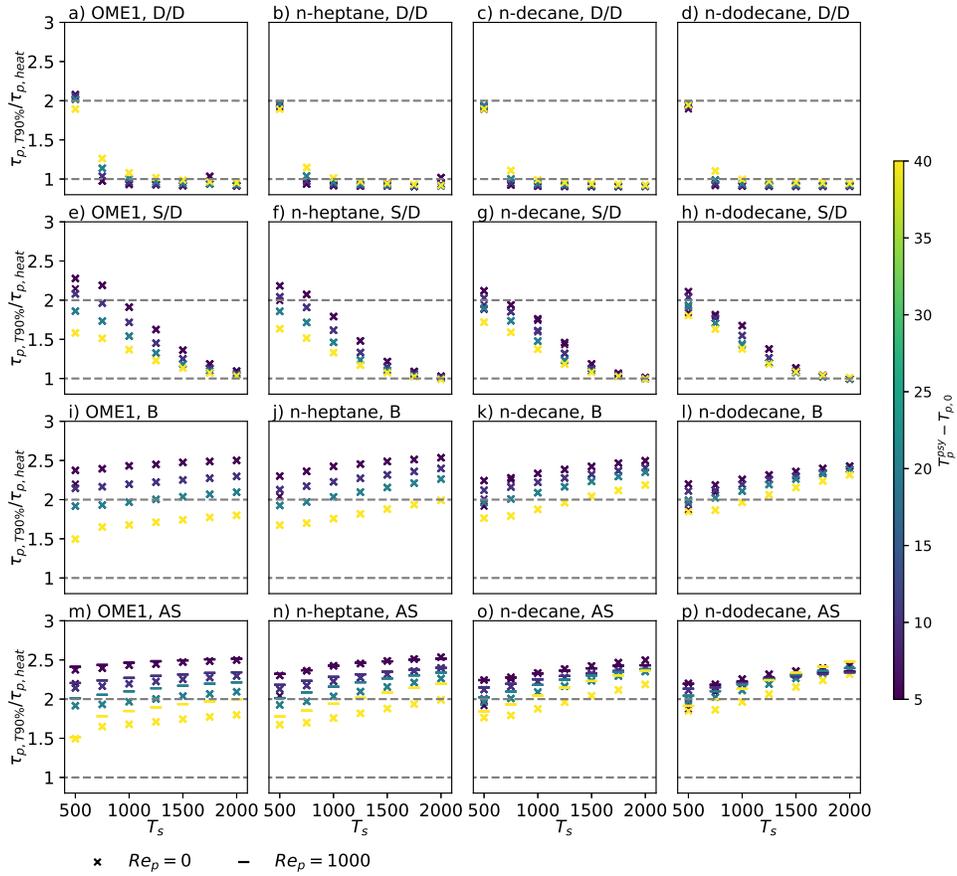}
    \caption{Comparison of the heat-up timescale estimate $\tau_{p,heat}$ and the time necessary to complete $90\%$ of the total temperature change in simulations. The ratio of the two time scales is assessed as function of the seen gas temperature under different initial temperatures marked by the color scheme, under different constant Reynolds numbers marked by the symbols, and with different initial droplet sizes. \mynormaltext{The droplet size is not indicated as there is no dependence, and similarly the Reynolds number is only indicated for the Abramzon-Sirignano model (AS).} }
    \label{fig:heatup_estimate}
\end{figure}

Figure~\ref{fig:heatup_estimate} shows the ratio of the simulated heat-up time scale and the estimate derived from the initial temperature slope: $\tau_{p,T90\%} / \tau_{p,heat}$\mynormaltext{, for the single droplet simulation cases of subsection~\ref{subsec:single_drop}.}
This ratio is displayed in a similar manner as in Fig.~\ref{fig:evap_estimate}.
\mynormaltext{The influence of initial droplet diameter is negligible on this property, thus the symbols are overlapping.
Similarly, this timescale ratio is not dependent on the Reynolds number except in the case of the Abramzon-Sirignano model (AS), thus the otherm models only display the $Re_p=0$ case. 
Two different Reynolds numbers are shown for AS, to illustrate the small influence of the Reynolds number on the heat up estimation.}

Taking Bird's correction (B) as reference, the $\tau_{p,T90\%} / \tau_{p,heat}$ ratio ranges between $1.5$ and $2.5$, indicating a behavior similar to the example shown in Fig.~\ref{fig:time_scale_illustrations}, where the droplet temperature smoothly transitions to the wet-bulb condition. 
There is no variation in terms of $d_{p,0}$ and $Re_p$, and even the seen temperature only has a weak effect on this ratio. 
At a given fuel, the ratio is varied the most by the initial temperature difference $T_{p}^{psy}-T_{p,0}$, indicating, that the temperature evolution during the heat-up period is not self-similar, but depends on the initial value. 
The dependence on the initial temperature diminishes as the volatility of the fuels decrease, i.e.: the variation is highest for OME1, and it diminishes almost completely for n-dodecane, especially under high seen gas temperatures. 
\mynormaltext{The high variability of the displayed ratio indicates, that the $\tau_{p,heat}$ estimate can only be used, to determine the order of magnitude of the heat-up period, but it is not accurate enough to define an exact relation.}

The different modeling strategies are also compared on Fig.~\ref{fig:heatup_estimate}.
The Abramzon-Sirignano model (AS) behaves almost identically to Bird's correction (B), except that the  $\tau_{p,T90\%} / \tau_{p,heat}$ ratio slightly increases with the Reynolds number, further comparison is provided below. 
In the case of the diffusion only model (D/D), the two regimes shown in Fig.~\ref{fig:validation} are clearly distinguishable. For $T_s = 500 \ \mathrm{K}$, this model is able to provide a stable solution, and consequently the ratio is approximately 2, since the droplet temperature reaches wet-bulb condition smoothly. In case of the rest of the seen temperatures, the diffusion only model (D/D) is unable to find stable solutions, and the droplet reaches the boiling point in approximately one $\tau_{p,heat}$ time. 
The classical model (S/D) shows very similar $\tau_{p,T90\%} / \tau_{p,heat}$ ratio to Bird's correction (B) at lower seen gas temperatures, but it transitions to fast heat-up as the seen gas temperature increases. This is due to the inconsistent consideration of Stefan flow in only the mass transfer, as this model needs orders of magnitudes higher Spalding mass transfer numbers than Bird's correction (B) to maintain the energy balance (see Fig.~\ref{fig:psych_wet_bulb} and Fig.~\ref{fig:psych_wet_bulb_bird}). 
Thus, in the transient cases of the present analysis, the latent heat of evaporation only starts to have a significant effect, once the droplet temperature is near the boiling point in case of high seen gas temperatures.

 \bibliographystyle{elsarticle-num} 
 \bibliography{main}

\begin{thebibliography}{10}
\expandafter\ifx\csname url\endcsname\relax
  \def\url#1{\texttt{#1}}\fi
\expandafter\ifx\csname urlprefix\endcsname\relax\def\urlprefix{URL }\fi
\expandafter\ifx\csname href\endcsname\relax
  \def\href#1#2{#2} \def\path#1{#1}\fi

\bibitem{sazhin2014droplets}
S.~Sazhin, Droplets and sprays, Vol. 345, Springer, 2014.

\bibitem{jenny2012modeling}
P.~Jenny, D.~Roekaerts, N.~Beishuizen, Modeling of turbulent dilute spray
  combustion, Progress in Energy and Combustion Science 38~(6) (2012) 846--887.

\bibitem{miller1998evaluation}
R.~Miller, K.~Harstad, J.~Bellan, Evaluation of equilibrium and non-equilibrium
  evaporation models for many-droplet gas-liquid flow simulations,
  International Journal of Multiphase Flow 24~(6) (1998) 1025--1055.

\bibitem{verdier2017experimental}
A.~Verdier, J.~M. Santiago, A.~Vandel, S.~Saengkaew, G.~Cabot, G.~Grehan,
  B.~Renou, Experimental study of local flame structures and fuel droplet
  properties of a spray jet flame, Proceedings of the Combustion Institute
  36~(2) (2017) 2595--2602.

\bibitem{both2017rans}
A.~Both, RANS-FGM simulation of n-heptane spray flame in OpenFOAM, 2017.

\bibitem{noh2018comparison}
D.~Noh, S.~Gallot-Lavall{\'e}e, W.~P. Jones, S.~Navarro-Martinez, Comparison of
  droplet evaporation models for a turbulent, non-swirling jet flame with a
  polydisperse droplet distribution, Combustion and Flame 194 (2018) 135--151.

\bibitem{sitte2019large}
M.~P. Sitte, E.~Mastorakos, Large eddy simulation of a spray jet flame using
  doubly conditional moment closure, Combustion and flame 199 (2019) 309--323.

\bibitem{ausilio2019study}
D.~Alessandro, I.~Stankovic, B.~Merci, Les study of a turbulent spray jet: mesh
  sensitivity, mesh-parcels interaction and injection methodology, Flow,
  Turbulence and Combustion 103~(2) (2019) 537--564.

\bibitem{chatelier2020large}
A.~Chatelier, B.~Fiorina, V.~Moureau, N.~Bertier, Large eddy simulation of a
  turbulent spray jet flame using filtered tabulated chemistry, Journal of
  Combustion 2020 (2020).

\bibitem{benajes2021analysis}
J.~Benajes, J.~M. Garc{\'\i}a-Oliver, J.~M. Pastor, I.~Olmeda, A.~Both,
  D.~Mira, Analysis of local extinction of a n-heptane spray flame using
  large-eddy simulation with tabulated chemistry, Combustion and Flame (2021)
  111730.

\bibitem{poinsot2005theoretical}
T.~Poinsot, D.~Veynante, Theoretical and numerical combustion, RT Edwards,
  Inc., 2005.

\bibitem{bird1960transport}
R.~B. Bird, W.~E. Stewart, E.~N. Lightfoot, Transport phenomena. 1960, 1960.

\bibitem{frossling1938uber}
N.~Fr\"{o}ssling, \"{U}ber die verdunstung fallernder tropfen, Gerlands
  Beitr\"{a}ge zur Geophysik 52 (1938) 170--216.

\bibitem{ranz1952evaporation}
W.~E. Ranz, W.~R. Marshall, Evaporation from drops: Part 1, Chem. eng. prog
  48~(3) (1952) 141--146.

\bibitem{yuen1976drag}
M.~C. Yuen, L.~W. Chen, On drag of evaporating liquid droplets, Combustion
  Science and Technology 14~(4-6) (1976) 147--154.
\newblock \href {https://doi.org/10.1080/00102207608547524}
  {\path{doi:10.1080/00102207608547524}}.

\bibitem{ebrahimian2011towards}
V.~Ebrahimian, C.~Habchi, Towards a predictive evaporation model for
  multi-component hydrocarbon droplets at all pressure conditions,
  International Journal of Heat and Mass Transfer 54~(15-16) (2011) 3552--3565.

\bibitem{daubert1985data}
T.~E. Daubert, R.~P. Danner, Data compilation tables of properties of pure
  compounds, Design Institute for Physical Property Data, American Institute of
  Chemical Engineers, 1985.

\bibitem{abramzon1989droplet}
B.~Abramzon, W.~A. Sirignano, Droplet vaporization model for spray combustion
  calculations, International journal of heat and mass transfer 32~(9) (1989)
  1605--1618.

\bibitem{gatley2005understanding}
D.~P. Gatley, Understanding psychrometrics, 2005.

\bibitem{godsave1953studies}
G.~Godsave, Studies of the combustion of drops in a fuel spray—the burning of
  single drops of fuel, in: Symposium (international) on combustion, Vol.~4,
  Elsevier, 1953, pp. 818--830.

\bibitem{chauveau2019analysis}
C.~Chauveau, M.~Birouk, F.~Halter, I.~G{\"o}kalp, An analysis of the droplet
  support fiber effect on the evaporation process, International Journal of
  Heat and Mass Transfer 128 (2019) 885--891.

\bibitem{depredurand2010heat}
V.~Depr{\'e}durand, G.~Castanet, F.~Lemoine, Heat and mass transfer in
  evaporating droplets in interaction: influence of the fuel, International
  journal of heat and mass transfer 53~(17-18) (2010) 3495--3502.

\bibitem{castanet2016evaporation}
G.~Castanet, L.~Perrin, O.~Caballina, F.~Lemoine, Evaporation of closely-spaced
  interacting droplets arranged in a single row, International Journal of Heat
  and Mass Transfer 93 (2016) 788--802.

\end{thebibliography}

\end{document}